\documentclass[superscriptaddress,twocolumn,10pt]{revtex4-2}

\usepackage{graphicx}
\usepackage{dcolumn}
\usepackage{bm}
\usepackage{todonotes} 
\usepackage{mathrsfs} 
\usepackage[version=4]{mhchem} 
\usepackage{siunitx} 
\usepackage{enumitem} 
\usepackage{color}
\newcommand{\figwidth}{0.85\linewidth}
\usepackage{float} 
\usepackage{url}

\begin{document}

\title{Elastic Properties of Confined Fluids from Molecular Modeling \\to Ultrasonic Experiments on Porous Solids}

\author{Christopher D. Dobrzanski}
\affiliation{Otto H. York Department of Chemical and Materials Engineering, New Jersey Institute of Technology, 323 Dr Martin Luther King Jr Blvd, Newark, NJ 07102, USA}
\author{Boris Gurevich}
\affiliation{Curtin University, Perth, Western Australia, Australia} %
\author{Gennady Y. Gor}
\email[Corresponding author, e-mail: ]{gor@njit.edu} 
\homepage[URL: ]{http://porousmaterials.net}
\affiliation{Otto H. York Department of Chemical and Materials Engineering, New Jersey Institute of Technology, 323 Dr Martin Luther King Jr Blvd, Newark, NJ 07102, USA}

\date{27 May 2021}

\begin{abstract}
Fluids confined in nanopores are ubiquitous in nature and technology. In recent years, the interest in confined fluids has grown, driven by research on unconventional hydrocarbon resources -- shale gas and shale oil, much of which are confined in nanopores. When fluids are confined in nanopores, many of their properties differ from those of the same fluid in the bulk. These properties include density, freezing point, transport coefficients, thermal expansion coefficient, and elastic properties. The elastic moduli of a fluid confined in the pores contribute to the overall elasticity of the fluid-saturated porous medium and determine the speed at which elastic waves traverse through the medium. Wave propagation in fluid-saturated porous media is pivotal for geophysics, as elastic waves are used for characterization of formations and rock samples. In this paper, we present a comprehensive review of experimental works on wave propagation in fluid-saturated nanoporous media, as well as theoretical works focused on calculation of compressibility of fluids in confinement. We discuss models that bridge the gap between experiments and theory, revealing a number of open questions that are both fundamental and applied in nature. While some results were demonstrated both experimentally and theoretically (e.g. the pressure dependence of compressibility of fluids), others were theoretically predicted, but not verified in experiments (e.g. linear scaling of modulus with the pore size). Therefore, there is a demand for the combined experimental-modeling studies on porous samples with various characteristic pore sizes. The extension of molecular simulation studies from simple model fluids to the more complex molecular fluids is another open area of practical interest. \\

\noindent \fbox{\begin{minipage}{0.775\linewidth}
This preprint is the draft version of the review article: \\
C.~D.~Dobrzanski, B.~Gurevich, G.~Y.~Gor \textit{Appl. Phys. Rev.} 8, 021317 (2021); \\
The final version is available at: \url{https://doi.org/10.1063/5.0024114
}
\end{minipage}}
\end{abstract}


\maketitle

\newpage

\tableofcontents

\newpage

\section{Introduction}
\label{sec:Intro}

Nanoporous materials, according to the IUPAC convention, are materials that have pore sizes below 100 nm \cite{Thommes2015}. Due to having such small pores and large surface areas, these materials are employed for many industrial applications including catalysis \cite{Weitkamp2000}, separation processes \cite{Li2009}, as adsorbents \cite{Peluso2019} or desiccants \cite{Byun2019}, as electrodes in energy storage \cite{Beguin2014}, and for methane storage \cite{Kumar2017}. Many of these processes focus on a specific desired effect on the fluids which are confined within the pores. Some geological materials, such as coal and shale are nanoporous and contain fluids within their pores \cite{Chalmers2012, Clarkson2013, Nie2015}. This spatial confinement and the interactions between the solid and fluid are known to induce changes to the solid structure of the nanoporous materials \cite{Gor2017review} as well as to the properties of the fluids confined within the nanopores \cite{Huber2015}. Changes of fluid properties due to confinement are widely discussed in the literature, they include density, melting point, diffusivity \cite{Gubbins2014, Huber2015, Barsotti2016}. The derivative thermodynamic properties, such as the thermal expansion coefficient \cite{Valenza2005, Garofalini2008, Xu2009}, are also altered by confinement, but have received much less attention. Derivative thermodynamic properties include the compressibility, which is the reciprocal to the bulk elastic modulus. This review focuses on the effects that confinement has on compressibility and other elastic properties. 

Elastic properties such as the bulk modulus, longitudinal modulus, and shear modulus are fundamental properties of a material and describe how a material responds to various mechanical loads. Knowledge of the elastic properties of confined fluids is important for probing the behavior and effectiveness of the fluids in various practical applications including high-pressure lubricants \cite{Martini2010}. Furthermore, the elastic moduli of a material also determine the speed at which elastic waves travel through the material. The quantitative understanding of elastic wave propagation in various media is of utmost importance for geophysics: seismic (tens of meters scale wavelength) and borehole-based sonic (cm-to-m scale) waves are used to characterize geological formations \textit{in situ}, and ultrasonic waves ($\mu$m-to-mm) are employed to characterize rock samples in the laboratory \cite{Sheriff1995}. Since most geological media are porous, and the pores are saturated with fluids (gas, water, brine, hydrocarbons, etc.), the elastic wave speed is controlled by the elastic moduli of both the solid and fluid components. If pores are macroscopic, the properties of the fluid in these pores are the same as in the bulk, but this is not necessarily true for fluids confined in nanopores. Unconventional hydrocarbon resources such as shale gas and shale oil are contained in the media that have substantial amount of nanopores \cite{Chalmers2012, Clarkson2013, Nie2015}. Thus, the recent progress in development of those resources motivates research in nanoporous media and confined fluids. Note that we are concerned exclusively with elastic properties of the confined fluids and their contribution to the elasticity of the nanoporous solids. Mechanical problems related to presence of organic matter and fracking are beyond the scope of our review.  

The main goals of this review are as follows:
\setlist{nolistsep}
\begin{enumerate}[noitemsep]
\item Overview the theoretical models employed to predict elastic properties of nanoconfined fluids (Section \ref{sec:Theory}). 
\item Describe the experimental methods for probing the elastic properties of fluid-saturated porous materials and relating them to the properties of nanoconfined fluids (Section \ref{sec:Expt}).
\item Analyze the theoretical predictions in the context of available experimental data (Section \ref{sec:Compare}).
\item Summarize the main experimental and theoretical findings and identify open questions related to elastic properties of nanoconfined fluids (Section \ref{sec:Summary}).
\end{enumerate}

\section{Theoretical Predictions}
\label{sec:Theory}

In the last two decades, molecular modeling has become a standard tool for studying physico-chemical properties of confined phases \cite{Gubbins2011}. Three molecular modeling techniques: Monte Carlo (MC) simulations, molecular dynamics (MD) simulations, and density functional theory (DFT), have been recently used for predicting elastic properties of confined fluids. This section summarizes theoretical results obtained using these methods. Additionally, we discuss the predictions of compressibility by equations of state for confined fluids. 

When a fluid is confined in the pore space of nanoporous solids, experiments can hardly probe the elastic properties of the fluid itself, they rather probe the solid-fluid composite (see detailed discussion in Section \ref{sec:Expt}). Molecular modeling, on contrary, can probe the fluid itself without considering the solid explicitly. Furthermore, since molecular simulations work well for small systems, it is even more natural to simulate the fluids alone, while considering the solid as just an external field. Thus, to our knowledge, all the theoretical works on elastic properties of fluids in nanopores reported the properties of the fluids themselves, rather than the properties of the solid-fluid composites probed in experiments. We will discuss the relation between the experimental data (for composites) and theoretical predictions (for fluid) in Section~\ref{sec:Compare}.

For most of the theoretical predictions, the main property of consideration is the isothermal compressibility $\beta_T$, the reciprocal of which is known as the isothermal bulk elastic modulus $K_T = \beta_T^{-1}$. For a macroscopic system, the isothermal compressibility is defined as
\begin{equation}
\label{beta-def1}
\beta_T \equiv - \frac{1}{V} \left( \frac{\partial V}{\partial P} \right)_{N,T} ,
\end{equation}
where $V$ is the system volume, $P$ is the fluid pressure, and $T$ is the absolute temperature. Here, following Refs.~\onlinecite{Landau5, Gor2015compr, Gor2016Tait}, we use the same definition of $\beta_T$ for the fluid confined in the pore. However, the definition Eq.~\ref{beta-def1} can be ambiguous because the fluid in confinement can be anisotropic. In this case it is described by the stress tensor~\cite{Irving1950, Schofield1982} (often referred to as the pressure tensor~\cite{Todd1995, Gray2011, Long2011, Long2012}). In addition to being anisotropic, the fluid is spatially inhomogeneous on the scale comparable to the nanopore size, and thus described in term of the local density. Similarly to the local density of the inhomogeneous fluid in the pore, other properties can be introduced in the local fashion, including the local compressibility. Several recent studies take these inhomogeneities into account; we discuss them in Section \ref{sec:Local}. However, here we start from the definition given by Eq.~\ref{beta-def1}, which provides a scalar property averaged over the pore volume. This overall compressibility of the fluid in the pore corresponds to the macroscopic average compressibility that can be extracted from experimental sound speed measurements on fluid-saturated porous samples overviewed in Section~\ref{sec:Expt}.

\subsection{Fluid Compressibility from an Adsorption Isotherm}
\label{sec:Isotherm}

When the pore space is filled by gas adsorption, the compressibility given by Eq.~\ref{beta-def1}, can be readily related to the adsorption isotherm -- amount adsorbed as a function of the pressure in the gas phase~\cite{Gor2014}. By neglecting the anisotropy of pressure and considering only a macroscopic average, the pressure $P$ in the pore, which is also known as the solvation pressure, can be determined from the grand thermodynamic potential $\Omega$~\cite{Ravikovitch2006, Gor2010}
\begin{equation}
\label{pressure1}
P = -\left( \frac{\partial \Omega}{\partial V} \right)_{\mu , T}.
\end{equation}
Also, the pressure in the pore $P$ is related to the chemical potential $\mu$ of the fluid via the Gibbs-Duhem equation
\begin{equation}
\label{gibbsduhem1}
{\rm d} P = n {\rm d} \mu
\end{equation}
where $n$ is the average particle density in the pore defined as $n \equiv N/V$. Assuming that the number of particles in the pore and the temperature are constant, Eq.~\ref{gibbsduhem1} can be used to rewrite Eq.~\ref{beta-def1} as
\begin{equation}
\label{beta-thermo1}
\beta_T = \frac{1}{n^2}\left( \frac{\partial n}{\partial \mu} \right)_{N, T}.
\end{equation}
Since, at constant temperature and when Eq.~\ref{gibbsduhem1} is valid, Eq.~\ref{beta-thermo1} is only a function of intensive variables (i.e., it does not depend on $N$ nor $V$), one can write
\begin{equation}
\label{partial-equality1}
\left( \frac{\partial n}{\partial \mu} \right)_{N, T} = \left( \frac{\partial n}{\partial \mu} \right)_{V, T}.
\end{equation}
This transformation is important because in the grand canonical ensemble, which is natural to model adsorption, the number of particles does indeed change while the volume of the system is kept constant. Thus, isothermal compressibility can be rewritten as
\begin{equation}
\label{beta-thermo2}
\beta_T = \frac{1}{n^2}\left( \frac{\partial n}{\partial \mu} \right)_{V, T}.
\end{equation}

\begin{figure}[H] \centering
\centering
\includegraphics[width=\figwidth]{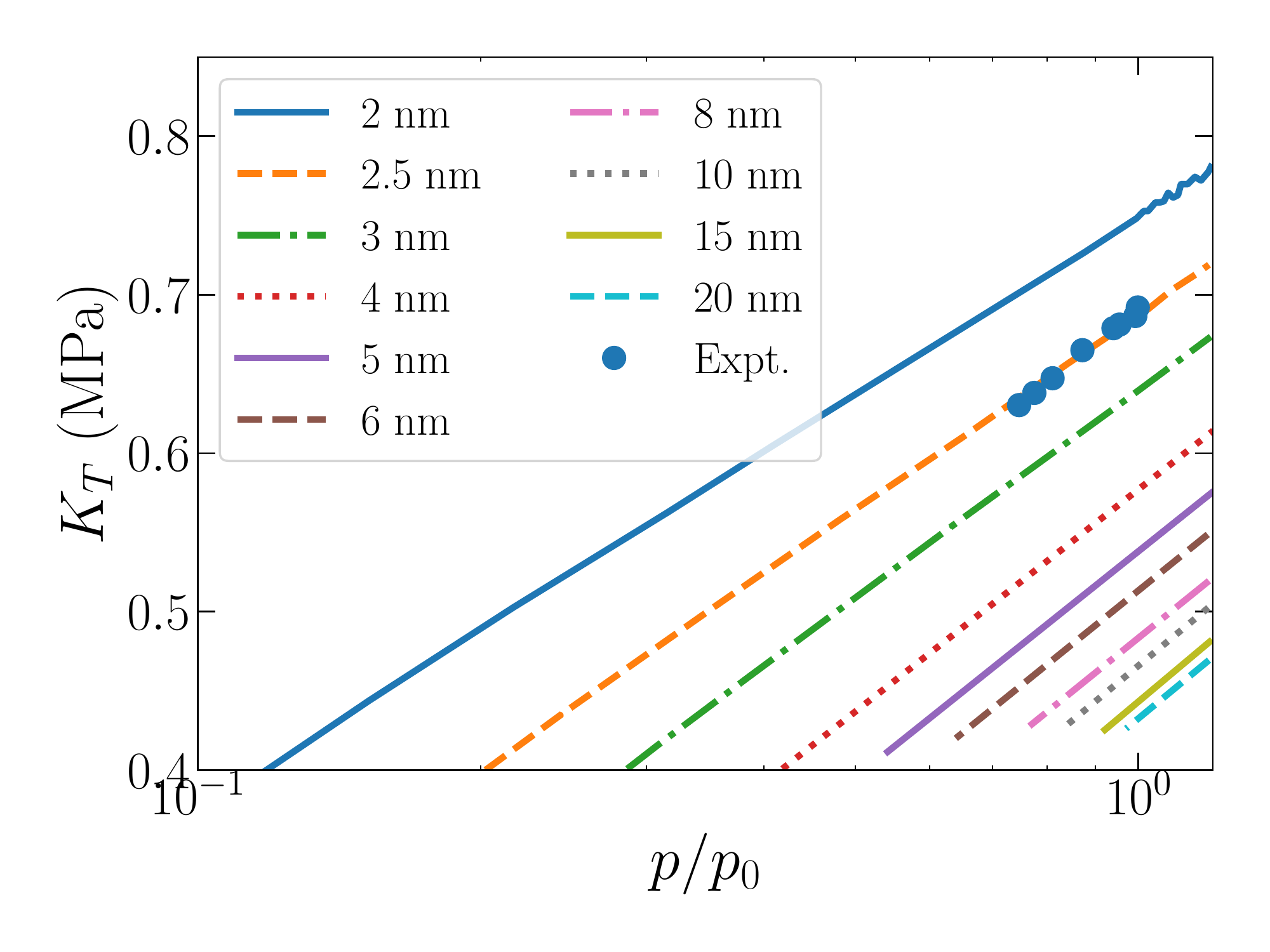}
\caption{Lines show the bulk modulus of liquid argon at $\SI{80}{K}$ confined in cylindrical nanopores as a function of relative gas pressure $p/p_0$ (after the capillary condensation) calculated using Eq.~\ref{beta-thermo-final1} from the QSDFT isotherms (lines). The markers show the values of isothermal modulus calculated from the ultrasonic data. Data from Refs.~\onlinecite{Schappert2014} and~\onlinecite{Gor2014}.}
\label{fig:Gor2014}
\end{figure}

For a single molecular species at equilibrium conditions, the chemical potential is related to the fugacity $f$ of the bulk fluid in equilibrium with the fluid in the pore by the relation
\begin{equation}
\label{chem-pot1}
\mu = k_{\rm{B}}T\ln(f/f_0) + \mu_0(T) ,
\end{equation}
where $f_0$ and $\mu_0(T)$ are the fugacity and chemical potential at saturation, respectively. Then Eq.~\ref{beta-thermo1} can be rewritten using Eqs.~\ref{partial-equality1} and \ref{chem-pot1} as~\cite{Gor2014}
\begin{equation}
\label{beta-thermo-final1}
\beta_T = \frac{1}{n^2} \frac{f/f_0}{k_{\rm{B}}T}\left( \frac{\partial n}{\partial (f/f_0)} \right)_{V, T}.
\end{equation}
Furthermore, when the vapor pressure is low (which is the case for argon at 80~K considered in Figure~\ref{fig:Gor2014}), the vapor can be considered an ideal gas, then the fugacity ratio $f/f_0$ can be replaced with the pressure ratio $p/p_0$, where $p_0$ is the vapor pressure at saturation.

Therefore, to calculate the compressibility of a confined fluid using the thermodynamic method, one only needs the density $n$ of the fluid in the pore as a function of the relative fugacity $f/f_0$, which is the adsorption isotherm. The derivative in Eq.~\ref{beta-thermo-final1} can be obtained from the slope of the isotherm. Fig.~\ref{fig:Gor2014} shows the bulk modulus $K_T = \beta_T^{-1}$ of confined liquid argon calculated using Eq.~\ref{beta-thermo-final1} from the theoretical isotherms generated using quenched solid density functional theory (QSDFT) \cite{Ravikovitch2006} for the fluid confined in pores of various size. Fig.~\ref{fig:Gor2014} compares the QSDFT prediction to the $K_T$ calculated from experimental ultrasonic data from Ref.~\onlinecite{Schappert2014}, showing qualitative agreement. This agreement is impressive given the approximate nature of Eq.~\ref{beta-thermo-final1}, based on the Gibbs-Duhem relation, which is strictly speaking only for the bulk system. A detailed discussion of comparison of theoretical prediction of confined fluids compressibility with experimental data from ultrasonic measurements is given in Section \ref{sec:Compare}.

\subsection{Compressibility from Monte Carlo and Molecular Dynamics Simulations}
\label{sec:MC}

Statistical mechanics provides a number of formulas based on fluctuations of various properties in statistical ensembles to calculate derivative properties (see e.g. Refs.~\onlinecite{Landau5, Allen2017}). Among different statistical mechanical ensembles and associated simulation techniques for molecular modeling, most hold the number of particles in the system constant. The grand canonical Monte Carlo (GCMC) \cite{Norman1969} algorithm is natural for modeling adsorption of fluids because it allows the number of particles in the pore (i.e., adsorbed) to change in accordance with the assigned chemical potential (or vapor pressure) of an external reservoir in equilibrium with the fluid in the pore, mimicking adsorption experiments. In this case, the isothermal compressibility of the fluid in the pore can be calculated from the fluctuations in the number of particles $N$ in the pore during GCMC simulations through the following relation 
\begin{equation}
\label{beta-fluct1}
\beta_T = \frac{V\langle \delta N^2\rangle}{k_{\rm{B}}T\langle N\rangle ^2} ,
\end{equation}
where $\langle \delta N^2\rangle$ is the variance of $N$ and $k_{\rm{B}}$ is the Boltzmann constant. Applying Eq.~\ref{beta-fluct1} to a small system requires that the fluctuation of $N$ obeys a Gaussian distribution~\cite{Landau5, Gor2015compr}. Thus, molecular simulation of a fluid in the pore performed in the grand canonical ensemble can provide data for calculation of $\beta_T$. 

A number of studies report the compressibility of confined fluids calculated using the GCMC simulation technique and applying Eq.~\ref{beta-fluct1} to the simulation data. Most of these works focus on the use of compressibility as a qualitative measure of a phase transition, in particular, on the phase transition of water in hydrophobic confinement. For example, Bratko et al.~\cite{Bratko2001} calculated the reduced isothermal compressibility $\beta_T^{\rm R} = { \beta_T k_{\rm B} T}/{V} = \langle \delta N^2 \rangle /\langle N \rangle ^2$ of a fluid between parallel plates with separation distances ranging between 1 and 6 nm. They found that the reduced compressibility enhances significantly as the separation distance decreases. They also found this enhancement of the reduced compressibility to be larger at lower values of vapor pressure for the same pore size. The follow-up studies exploring the effects of an electric field on water in hydrophobic confinement also employed isothermal compressibility calculated based on the fluctuations of number of particles as a measure for vapor-liquid phase transition \cite{Vaitheeswaran2005, Bratko2007}. Calculating the reduced compressibility avoids the questionable nature of defining the volume $V$ used in calculating the compressibility of the confined fluid \cite{Keffer1996, Rasmussen2010}, which makes it convenient for purely theoretical qualitative analysis. At the same time, the use of reduced compressibility does not allow a comparison to experimental data (which accordingly was not attempted in Ref.~\onlinecite{Bratko2007}).

\begin{figure}[H] \centering
\centering
\includegraphics[width=\figwidth]{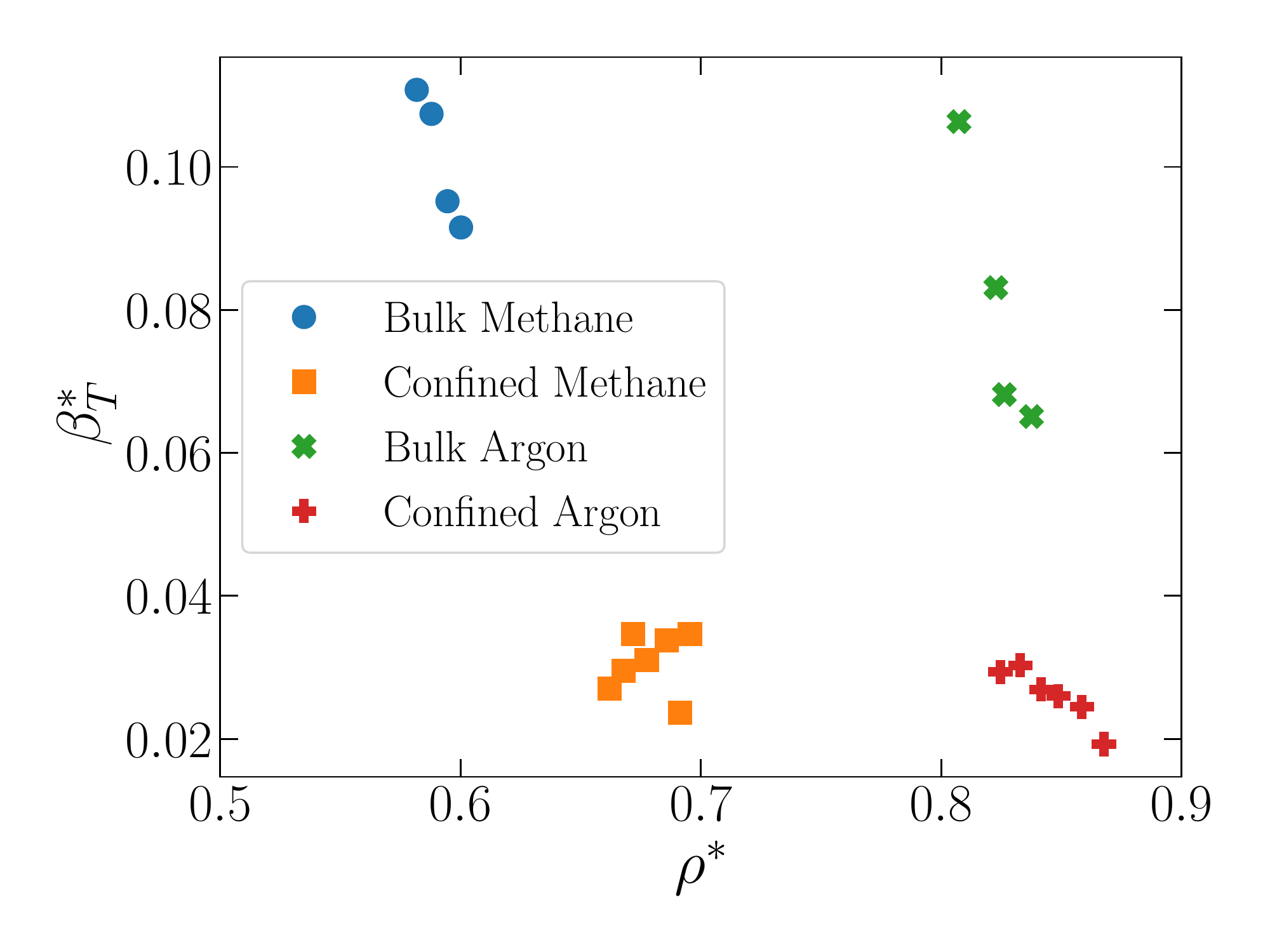}
\caption{Reduced isothermal compressibility $\beta_T^* = \beta_T \varepsilon / \sigma^3$ of methane and argon where $\varepsilon$ and $\sigma$ are the Lennard-Jones (LJ) parameters for the fluid plotted versus reduced density $\rho^* = \rho \sigma^3$. The confined argon shows lower compressibility and slightly higher density compared to the fluid in bulk. Data from Ref.~\onlinecite{Coasne2009}.}
\label{fig:Coasne2009}
\end{figure}

Compressibility of a liquid typically changes significantly in the course of the phase transition, e.g. freezing. Hence, Coasne et al.~\cite{Coasne2009} calculated compressibilities in order to understand the freezing behavior of fluids in confinement and how it depends on pressure (See also Section \ref{sec:Shear}). They utilized Eq.~\ref{beta-fluct1} to calculate the compressibility of argon and methane confined in graphene slit-like pores; the widths of the pores were twice the molecular diameter of the fluid. They found that the compressibility of the confined fluid was about 1/2 and 1/3 of the bulk fluid values of argon and methane, respectively. The freezing temperatures of bulk fluids typically have weak dependence on pressure due to low compressibility; however Coasne et al. found a significant dependence for the confined fluid. They cited this lower compressibility of the confined fluid as evidence that the significant pressure dependence of the freezing temperature is unrelated to the compressibility. Their data are shown in Figure \ref{fig:Coasne2009}. Recently, GCMC and Eq.~\ref{beta-fluct1} were utilized to calculate compressibility of confined liquid argon and nitrogen in silica pores, in order to compare the predictions to the values measured in ultrasonic experiments. These results are discussed in detail in Section~\ref{sec:Compare}.

Alternatively to calculating compressibility from the fluctuation of number of molecules in the grand canonical ensemble (Eq.~\ref{beta-fluct1}), one can use the volume fluctuations in the isothermal-isobaric ensemble:
\begin{equation}
\label{beta-fluct2}
\beta_T = \frac{\langle \delta V^2\rangle}{k_{\rm B}T \langle V \rangle} .
\end{equation}
This approach was utilized by Strekalova et al.~\cite{Strekalova2011, Strekalova2012} for studying water in hydrophobic confinement around nanoparticles. Performing the MC simulations, they found that there is a first-order liquid-liquid phase transition associated with an over 90\% decrease in the compressibility in the region of the phase transition. They found that a nanoparticle concentration of just 2.4\% is enough to prevent the liquid-liquid phase transition at pressures above 0.16 GPa. 

Another fluctuation formula utilized recently for calculation of a confined fluid compressibility is based on the simulations in canonical ensemble~\cite{Allen2017}
\begin{equation}
\label{beta-NVT}
\begin{gathered}
\beta_T^{-1} = K_T \\ 
= \frac{1}{V}\left(Nk_{\rm B}T + \langle \mathscr{W} \rangle_{NVT} + \langle \mathscr{X} \rangle_{NVT} - \frac{\langle \delta \mathscr{W}^2 \rangle_{NVT}}{k_{\rm B}T}\right),
\end{gathered}
\end{equation} 
where $\mathscr{W}$ is the internal virial, $\langle \delta \mathscr{W}^2 \rangle_{NVT}$ is the variance of the internal virial, and $\mathscr{X}$ is a hypervirial function. Corrente et al. utilized Eq.~\ref{beta-NVT} for calculating compressibility of methane confined in carbon nanopores, which was to model the natural gas found in coal and shale systems~\cite{Corrente2020}. They performed simulations on slit pores of widths ranging from 2 to 9 nm using GCMC and molecular dynamics (MD) simulations in $NVT$ ensemble. The results of the calculations using Eqs.~\ref{beta-fluct1} from GCMC and \ref{beta-NVT} from MD appeared fully consistent with each other.

\begin{figure}[H] \centering
\includegraphics[width=\figwidth]{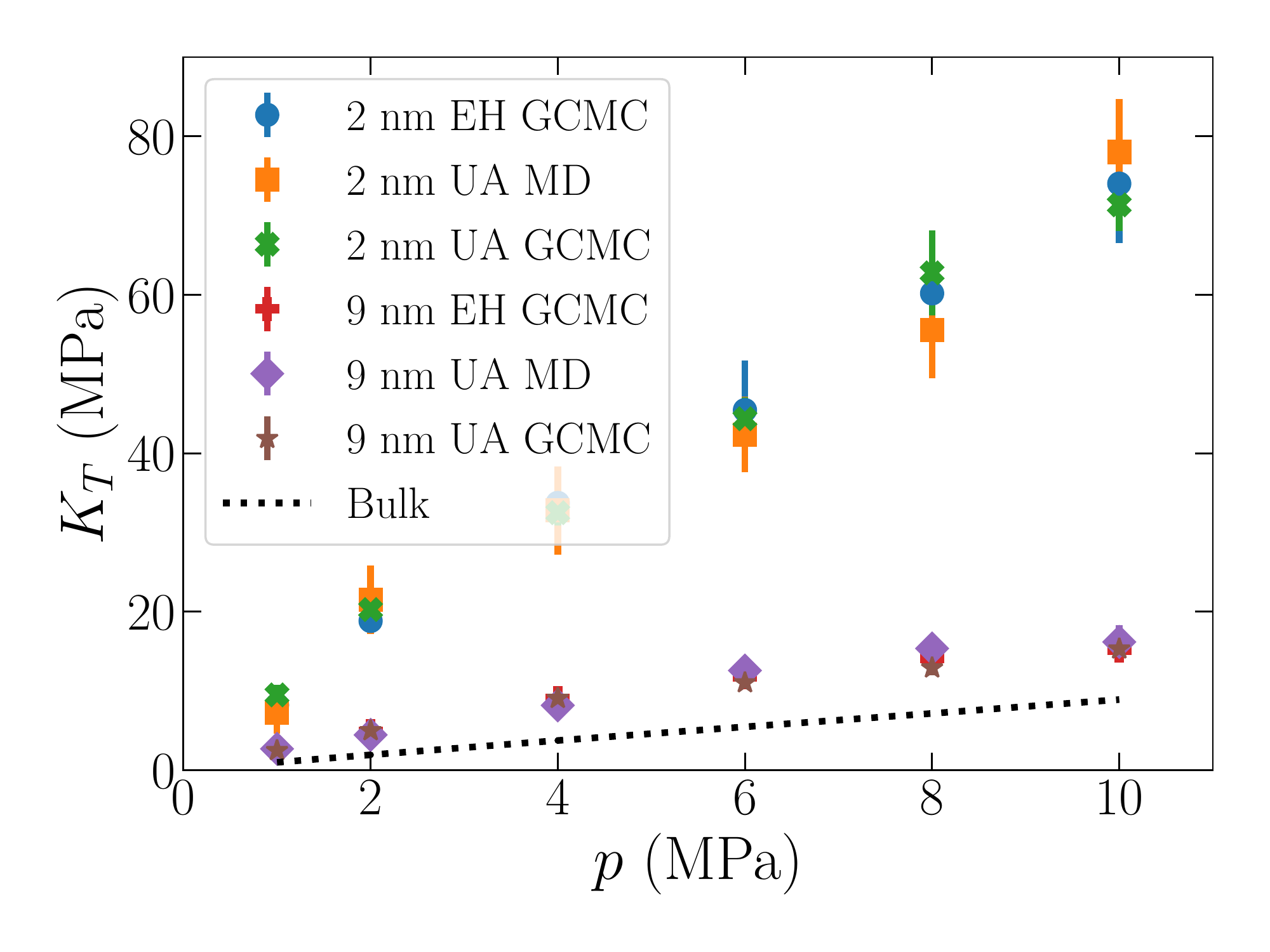}
\caption{Isothermal modulus of methane confined in carbon pores at 298 K as a function of bulk reservoir pressure from GCMC and MD simulations. The points represent calculations done in the 2 and 9 nm pores using either grand canonical Monte Carlo (GCMC) or molecular dynamics (MD) simulations. Methane was modeled using an explicit hydrogen (EH) model, as well as a more convenient united-atom (UA) model, which had good agreement. The dotted line represents the bulk methane modulus. The calculations from GCMC are based on the fluctuation of number of particles (reciprocal of Eq. \ref{beta-fluct1}). The MD calculations are done in \textit{NVT} ensemble where the fluid modulus is calculated using Eq. \ref{beta-NVT}. Data from Ref.~\onlinecite{Corrente2020}.}
\label{fig:Corrente} 
\end{figure}

Figure \ref{fig:Corrente} shows the data on elastic modulus of methane confined in carbon pores of two different pore sizes as a function of pressure. Different lines correspond to the GCMC and MD methods used for calculations and also to two different models for methane -- an explicit-hydrogen (EH) model where the all the atoms of the molecule are explicitly modeled in the simulation, and the united-atom (UA) model where the methane intermolecular interactions are approximated with a Lennard-Jones (LJ) potential from a single site for each molecule. Simulations showed a higher modulus (lower compressibility) compared to bulk value, and that the modulus has a monotonic increase with increased pressure. Such substantial increase of elastic modulus of confined methane over the bulk value suggests that it can affect the other properties, in particular the speeds of wave propagation in nanoporous solids saturated with methane. 

In addition to various fluctuation formulas (Eqs.~\ref{beta-fluct1}, \ref{beta-fluct2}, and \ref{beta-NVT}), the compressibility (or modulus) of confined fluid can be calculated using molecular dynamics by direct simulation of the fluid compression. This straightforward approach was used by Martini and Vadakkepatt to calculate the modulus of a thin lubricant film behavior in a slit pore~\cite{Martini2010}. They modeled hexadecane fluid confined in 5 nm wide alumina slit pores at different temperatures (300, 350, 400 K) using MD simulation. They applied a small change in pressure via compressive load onto one of the pore walls while fixing the other and measured the resulting volume change. The changes in pressure and volume were used to calculate the compressibility via the definition Eq.~\ref{beta-def1}. The resulting modulus appeared somewhat lower than the modulus for the same fluid in bulk, which likely suggests a somewhat solvophobic confinement.

\subsection{Pressure-Modulus Relation}
\label{sec:Pressure-Modulus}

Several theoretical works explored the relation between the pressure in the confined fluid and its compressibility. When studying the pressure dependence, the bulk modulus is more natural to use than the compressibility, because for bulk fluids (and solids), the modulus is related to pressure with a simple linear relation, known as Tait-Murnaghan equation~\cite{Murnaghan1944, Birch1952}:
\begin{equation}
\label{Tait}
K(P) = K(P_0) + \alpha (P-P_0),
\end{equation}
where the dimensionless constant $\alpha$ is the slope of the observed linear dependence. Eq.~\ref{Tait} is simply the first two terms of the Taylor series of $K(P)$, consequently it is rather general and does not depend on whether applied to bulk or confined fluid. 

Qualitatively, the relation between the pressure and elastic properties of a confined fluid is transparent: the attractive solid-fluid interactions densify the fluid near the pore walls, making it effectively compressed~\cite{Aranovich2003}. This compression can be described in terms of the solvation pressure $P$, reaching tens or hundreds of MPa; the same pressure which is the driving force for adsorption-induced deformation (see Section \ref{sec:Deformation}). Compressed fluid thus becomes stiffer -- the modulus increases with the pressure. The molecular dynamics data for hexadecane confined in 5 nm wide alumina slit pores at different temperatures showed a nearly linear dependence of the modulus on pressure for pressures up to $\SI{5}{GPa}$ \cite{Martini2010}. Note that the resulting curves reported by the authors were only slightly deviating from the bulk. 

\begin{figure}[H] \centering
\includegraphics[width=\figwidth]{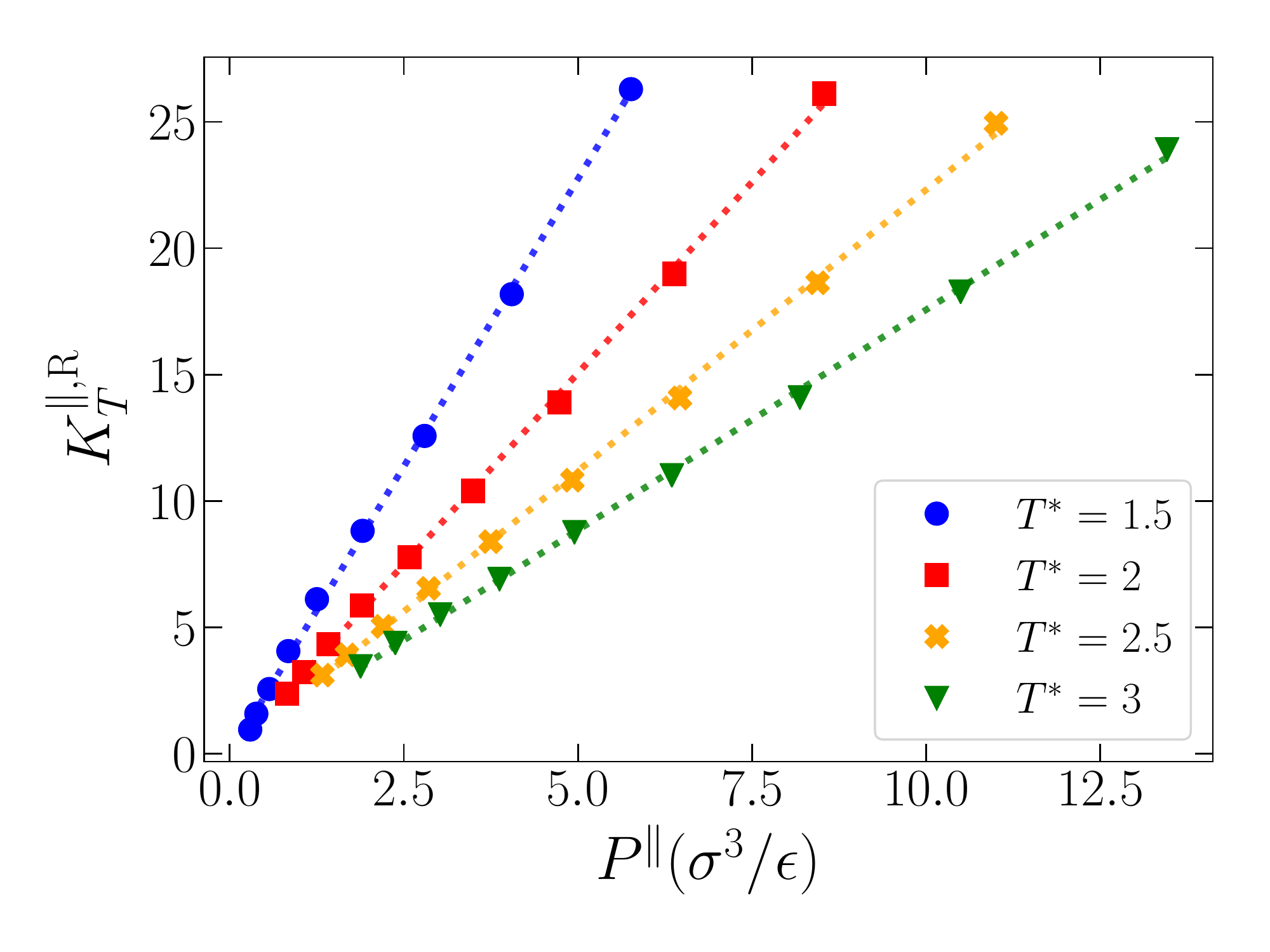} \\
\includegraphics[width=\figwidth]{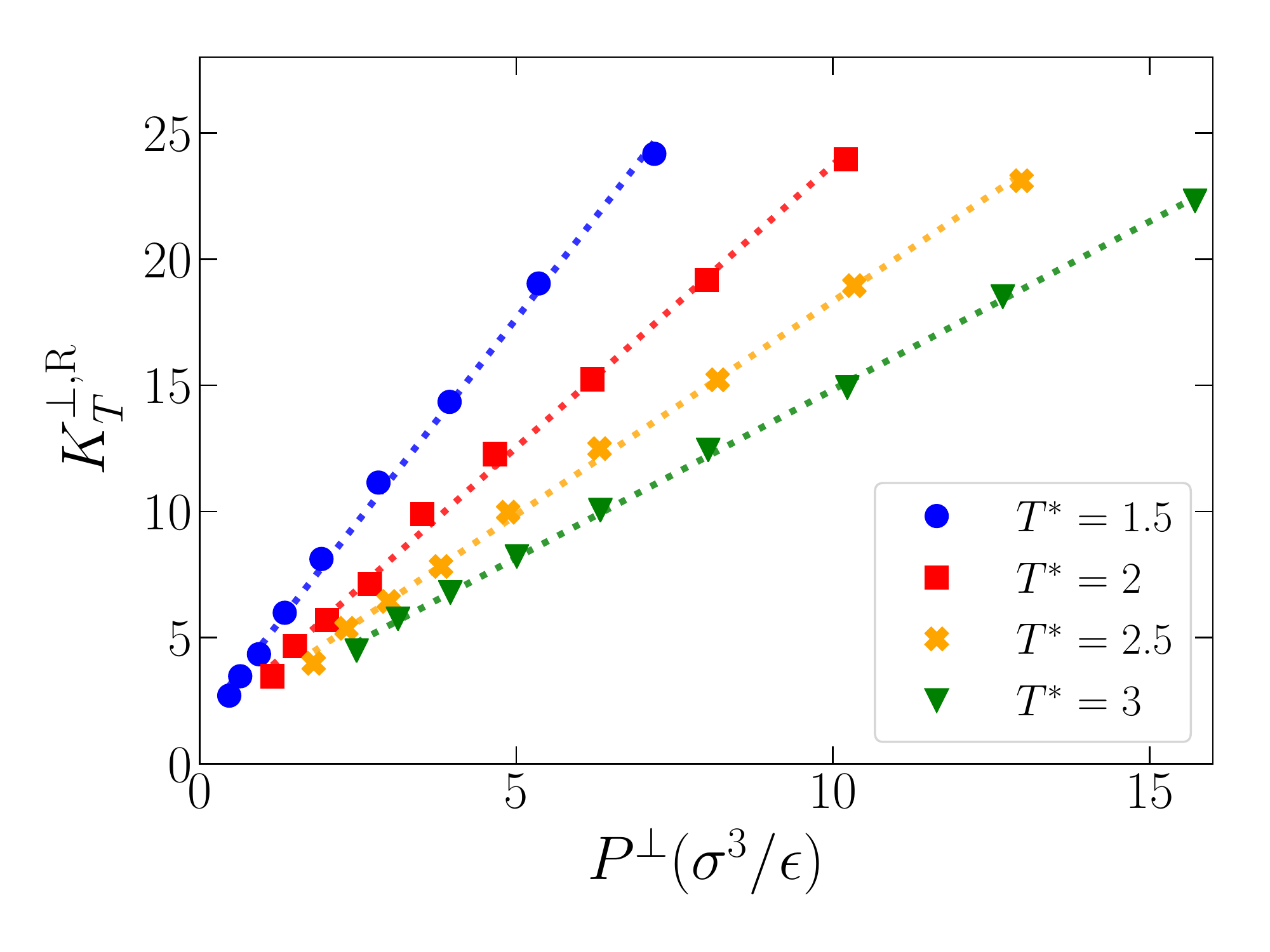}
\caption{Reduced bulk modulus versus reduced pressure for the LJ fluid confined in a slit-like pore of width 4$\sigma$ at LJ reduced temperatures of $T^* =$ 1.5, 2, 2.5, and 3. Top: the lateral component of bulk modulus versus lateral component of pressure. Bottom: the normal component of bulk modulus versus normal component of pressure. Dotted lines are linear fits of the points corresponding to the same color. Data from Ref.~\onlinecite{Keshavarzi2016}.}
\label{fig:Keshavarzi} 
\end{figure}

The pressure-modulus relation was recently studied for a LJ fluid confined in a LJ slit pore using the classical density functional theory (DFT)~\cite{Keshavarzi2016}. Keshavarzi et al. considered pores of widths between 2 to 8 multiples of $\sigma$ (LJ distance unit) and at the reduced temperatures between 1.5 and 3, and calculated the average isothermal modulus from the average density of the fluid in the pore $n$ as 
\begin{equation}
\label{K-n}
K_T = n \left( \frac{\partial P}{\partial n} \right)_{N,T}. 
\end{equation}
Taking into account the anisotropy of the fluid, they introduced the two moduli: normal $K_T^{\perp}$ and lateral $K_T^{\parallel}$ corresponding respectively to $P = P^{\perp}$ and $P = P^{\parallel}$ in Eq.~\ref{K-n}. They presented the resulting moduli as functions of corresponding pressures obtaining in both cases linear relations for each of the temperatures. This suggests that the confined fluid modulus dependence on pressure, similarly to the modulus of a bulk fluid, can be described by the Tait-Murnaghan equation. The data from Ref.~\onlinecite{Keshavarzi2016} are shown in Fig.~\ref{fig:Keshavarzi}; it is important to note that these data were reported in the format of reduced modulus $K^{\rm R}_T = K_T n/T$.

Another recent work used GCMC simulations to calculate the isothermal modulus of argon fluid confined in spherical nanopores, specifically focusing on the modulus-pressure relation~\cite{Gor2016Tait}. The results were consistent with that of Ref.~\onlinecite{Keshavarzi2016}, suggesting a linear Tait-Murnaghan relation holds between $K_T$ and the Laplace pressure (calculated simply from the chemical potential, Eq.~\ref{chem-pot1}). Gor et al.~\cite{Gor2016Tait} also varied the solid-fluid interaction strength to show how it influences the elastic modulus, finding that the higher interaction strengths were associated with higher moduli. Also, the calculated slope $\alpha$ in Eq.~\ref{Tait} for the confined fluid was found to match the slope for the fluid in bulk, as long as the interactions were not solvophobic~\cite{Gor2016Tait}. Interestingly, if Keshavarzi et al.~\cite{Keshavarzi2016} used not the reduced modulus, but reported the modulus as calculated by definition (Eq.~\ref{K-n}), their data would have shown nearly the same slope for all of their lines at different temperatures. Importantly, a recent experimental work by Schappert and Pelster reported that the slope of the proportionality constant $\alpha$ for confined argon is independent of the temperature~\cite{Schappert2018liquid}.

\subsection{Local Elastic Properties}
\label{sec:Local}

\begin{figure}[H] \centering
\centering
\includegraphics[width=\figwidth]{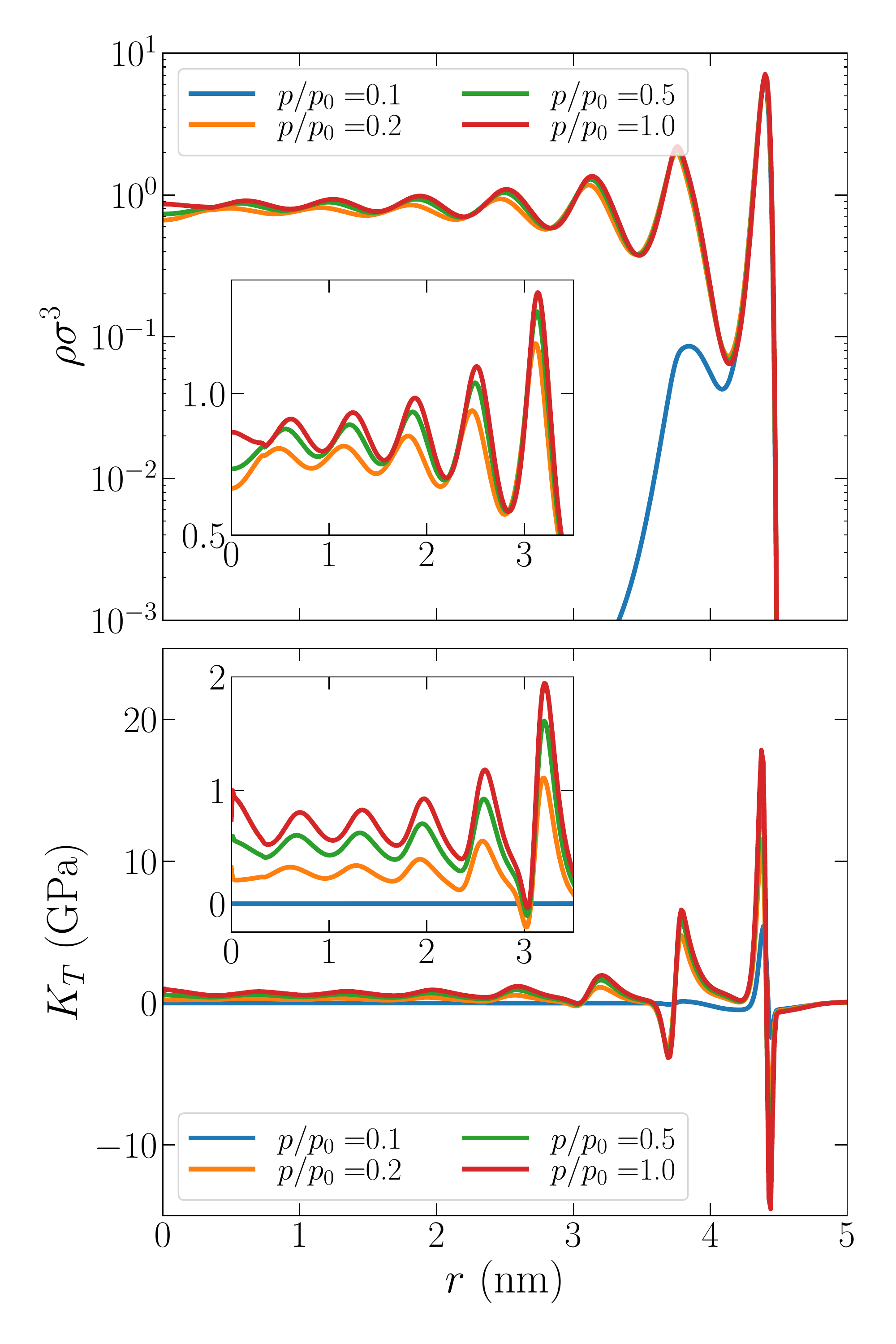} 
\caption{Fluid density profiles (top) and local modulus (bottom) for argon confined in a 5 nm spherical pore at $\SI{87.3}{K}$ and relative pressures $p/p_0 = $ 0.1, 0.2, 0.5, and 1.0. The insets show the variation of data further from the pore wall. Data from Ref.~\onlinecite{Sun2019density}.}
\label{fig:Sun-profiles}
\end{figure}

The density of fluids confined in nanopores is spatially dependent, with local maxima near the pore wall in the case of solvophilic confinement, and local minima in the case of solvophobic, e.g., the upper panel of Fig.~\ref{fig:Sun-profiles} shows the densities of LJ fluid confined in a spherical pore from Ref.~\onlinecite{Sun2019}. These inhomogeneities allow one to introduce local thermodynamic properties, such as a local pressure tensor~\cite{Long2013}. Similarly, the derivative thermodynamic properties, and in particular, the local fluid compressibility, can be introduced, as was done in several works within the last decade.

The local compressibility of a confined fluid can be calculated based on the elastic constant tensor components in k-space from an assumed linear relation between components of the stress rate and the strain rate~\cite{Schofield1966}. Rickman used this approach to determine local compressibility of LJ fluid confined in slit-shaped pores in Monte-Carlo simulations and related them to the fluid structure~\cite{Rickman2012}. He reported correlations of the local compressibility with the local density and the strength of fluid-wall interactions.

A different approach has been taken by Evans and coworkers~\cite{Evans2015, Evans2015PRL}: they defined the local compressibility using the density and chemical potential as~\cite{Tarazona1982} $\left( \frac{ \partial \rho(z)}{ \partial \mu } \right)_T$, where $z$ is the spatial coordinate. This allowed investigating the compressibility as a function of distance to the adsorbent wall. They performed DFT calculations of fluid near a single wall and of fluid confined between two walls. They found similar effects on their local compressibility in both cases, indicating confinement effects are largely due to the proximity of the fluid to the surface. They compared how different fluid-wall interactions affected the local compressibility and found that solvophobicity has a larger effect on the compressibility than on the density of the fluid, demonstrating that compressibility can be a good indicator of the solvophobicity of a surface~\cite{Evans2015, Evans2015PRL}. Later, Evans et al.~\cite{Evans2017} extended this method for GCMC simulations, which were found to be consistent with their DFT calculations. 

Application of DFT for calculation of local elastic properties was further used by Sun et al. who modeled argon in slit and later in spherical pores~\cite{Sun2014, Sun2019confinement, Sun2019density}. They formed the expressions for elastic moduli based directly on Hooke's law. One can relate the elastic modulus to changes in the stress tensor $\mathbf{\Pi}$ before and after deformation and the strain tensors $\mathbf{T}$. The change in the stress tensor is \cite{Fisher1964}
\begin{equation}
\label{Pi}
\mathbf{\tilde{\Pi}} - \mathbf{\Pi} = G_T (\mathbf{T}_{\alpha \beta} + \mathbf{T}_{\beta \alpha}) + \left( K_T - \frac{2}{3} G_T \right) \mathbf{T}_{\alpha \alpha} ,
\end{equation}
where $G_T$ and $K_T$ are the isothermal shear and bulk moduli, respectively. The stress tensor can be obtained from the Irving-Kirkwood expression~\cite{Irving1950}
\begin{equation}
\label{IK}
\begin{gathered}
\mathbf{\Pi} = -k_{\rm{B}} T \rho (\mathbf{r}) \mathbf{I} + \frac{1}{2} \int {\rm d} \mathbf{r}_{12} \frac{\mathbf{r}_{12} \mathbf{r}_{12}}{r_{12}} U ' (r_{12}) \\
\times \int_0^1 {\rm d} \xi \rho ^{(2)} (\mathbf{r} - \xi \mathbf{r}_{12}, \mathbf{r} - \xi \mathbf{r}_{12} + \mathbf{r}_{12}) ,
\end{gathered}
\end{equation}
where $\mathbf{r}_{12} = \mathbf{r}_2 - \mathbf{r}_1$, $r_{12} = | \mathbf{r}_{12} |$, $\rho (\mathbf{r})$ and $\rho^{(2)} (\mathbf{r}_1,\mathbf{r}_2)$ are the singlet and doublet pair density functions, respectively, $\mathbf{I}$ is the unit tensor, $U (r_{12})$ is the pair potential, and $\xi \in (0,1)$ is a constant. Using Eqs.~\ref{Pi}, \ref{IK} Sun et al. calculated the isothermal shear and bulk moduli as
\begin{equation}
G_T(\mathbf{r}) = k_{\rm{B}} T \rho (\mathbf{r}) + \frac{4}{15} {I}_1(\mathbf{r}) + \frac{1}{15} {I}_2(\mathbf{r}) 
\end{equation}
and
\begin{equation}
K_T(\mathbf{r}) = \frac{5}{3} k_{\rm{B}} T \rho (\mathbf{r}) - \frac{2}{9} {I}_1(\mathbf{r}) + \frac{1}{9} {I}_2(\mathbf{r}),
\end{equation}
where ${I}_1$ and ${I}_2$ are auxiliary integrals involving the pair density function, the details of which are in Refs.~\onlinecite{Fisher1964, Sun2014, Sun2019confinement, Sun2019density}.

\begin{figure}[H] \centering
\centering
\includegraphics[width=\figwidth]{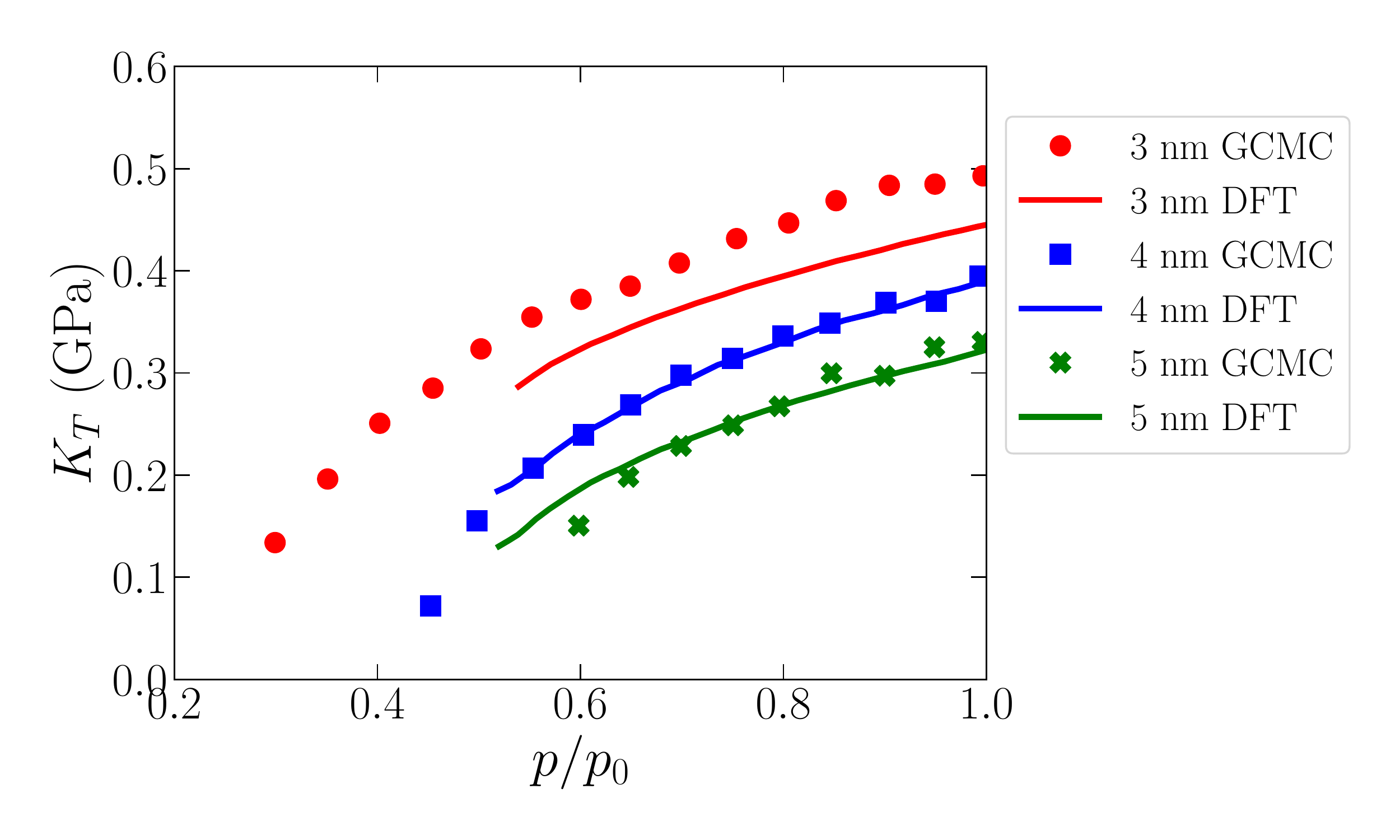}
\caption{Isothermal bulk modulus of confined argon at $T = \SI{119.6}{K}$ as a function of relative vapor pressure. The diameters of the spherical pores are 3, 4, and 5 nm. The markers are calculations based on GCMC simulations using Eq.~\ref{beta-fluct1}, and the solid lines represent density functional theory calculations. Data from Refs.~\onlinecite{Dobrzanski2018} and \onlinecite{Sun2019density}.}
\label{fig:Sun2019modulus}
\end{figure}

Finally, Sun et al.~\cite{Sun2019density} obtained an average of this modulus in the pore over the width of the pore $d$
\begin{equation}
\label{K-average}
\overline{K}_T = \frac{2}{d} \int_0^{d/2} K_T(r) {\rm d} r.
\end{equation}
They found that the elastic modulus has large deviations in the pore from the average value and can have large negative spikes. The negative modulus has been found to relate to the gas-liquid or liquid-solid transitions, which can be stabilized by confinement in nanopores \cite{Sun2019density}. The calculated average value of the isothermal modulus is consistent with other similar theoretical predictions, and in particular with the data from Dobrzanski et al~\cite{Dobrzanski2018} obtained for argon in silica pores by GCMC using Eq.~\ref{beta-fluct1} -- this comparison is shown in Fig.~\ref{fig:Sun2019modulus}. Of note, however, Sun et al. took the spatial average over the pore radius (Eq.~\ref{K-average}) rather than the pore volume for the spherical pore. The approach proposed by Sun et al. can be used further e.g. to calculate the modulus of adsorbed film, instead of the modulus of the fluid averaged over the entire pore. 

\subsection{Compressibility from Equations of State for Confined Fluids}
\label{sec:EOS}

Molecular simulations are powerful tools in modeling the behaviors and properties of materials down to the atomic level.
They enable modeling the confinement effects on elastic properties. Simulations have the potential to calculate the elastic properties of any system under any possible condition including temperature and pressure. However, in order to do so, the calculations would have to be carried out for each system, under each condition, and at each pore size, which would be computationally expensive, especially for dense fluids at low temperature \cite{Dobrzanski2018}. A more practical approach for predicting thermodynamic properties is based on equations of state (EOS). To accurately model the fluids in nanopores, the EOS needs to be developed with the effects of confinement in mind. Recently, there have been a number of attempts to model fluids under confinement using an EOS \cite{Schoen1998, Truskett2001, Giaya2002, Zarragoicoechea2002, Zarragoicoechea2004, Zhu1999, Alharthy2013, Teklu2014, Tan2015EOS1, Tan2015EOS2, Dong2016, Yang2018, Travalloni2010, Travalloni2010critical, Travalloni2014, Barbosa2016, Barbosa2018, Barbosa2019}. However, none of those works have been developed for the elastic properties.

Dobrzanski et al. explored the possibility of an EOS for confined fluids to predict the compressibility of the fluid \cite{Dobrzanski2020}. They used the generalized van~der~Waals (vdW) EOS developed by Travalloni et al. \cite{Travalloni2010, Travalloni2010critical} for square-well fluid confined in a cylindrical pore:
\begin{equation}
\label{eq:p}
\begin{gathered}
P = \frac{R_{\rm g} T}{v-b_{\rm{p}}} - \frac{a_{\rm{p}}}{v^2} - \theta \frac{b_{\rm{p}}}{v^2} \left( 1-\frac{b_{\rm{p}}}{v} \right) ^{\theta-1} (1-F_{\rm{pr}}) \\
\left[ R_{\rm g} T \left(1- \exp \left( - \frac{N_{\rm{A}} \epsilon_{\rm{p}}}{R_{\rm g} T} \right) \right) - N_{\rm{A}} \epsilon_{\rm{p}} \right] ,
\end{gathered}
\end{equation}
where $v$ is the molar volume, $N_{\rm A}$ is Avogadro's number, $\epsilon_{\rm p}$ is the energy parameter of the fluid-wall interaction, and $a_{\rm{p}}$ and $b_{\rm{p}}$ are the vdW EOS parameters modified by confinement in a pore of radius $r_{\rm p}$. The geometric function $F_{\rm pr}$ is the fraction of the confined fluid molecules within the square-well region of the interaction with the pore wall for a randomly distributed fluid. The parameter $\theta$ is the geometric parameter, related to the pore size, and the linear parameters of the interatomic potentials. 

\begin{figure}[H] \centering
\centering
\includegraphics[width=\figwidth]{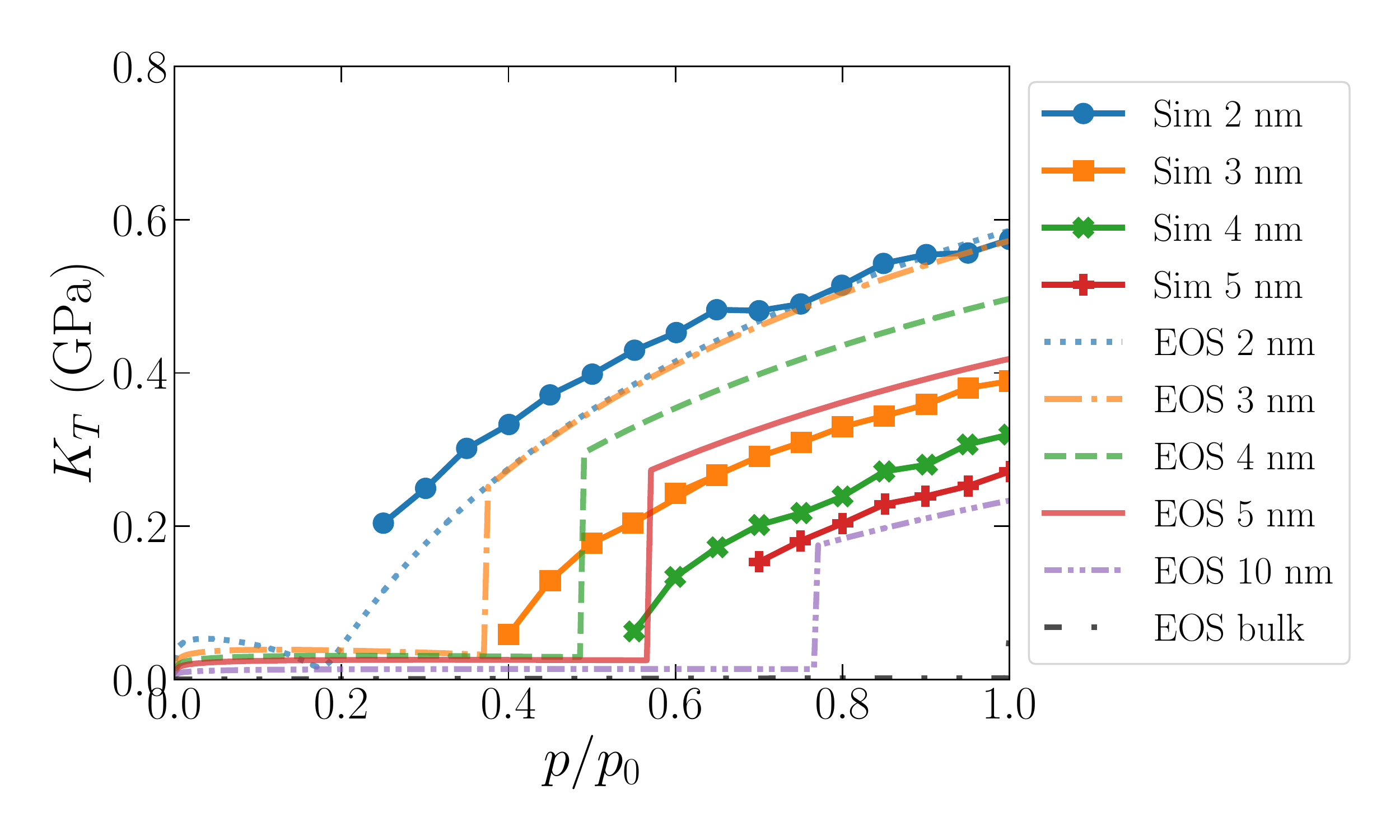}
\caption{Isothermal elastic modulus of argon confined in silica nanopores at 119.6 K calculated using GCMC simulation and the modulus derived from the EOS of Travalloni et al. \cite{Travalloni2010}. The plot shows the EOS can give the same behavior predicted from the simulations at different pressures and pore sizes. Data from Ref.~\onlinecite{Dobrzanski2020}.}
\label{fig:eosfigkiso}
\end{figure}

Eq.~\ref{eq:p} has been shown to be able to model fluid adsorption in nanopores \cite{Travalloni2010}. It is convenient because it has only two fitting parameters related to the solid-fluid interaction strength $\epsilon_{\rm{p}}$ and to the width of the fluid-wall interaction well $\delta_{\rm p}$. Dobrzanski et al. used this formalism and derived the following analytical expression for the isothermal elastic modulus of the confined fluid,
\begin{widetext}
\begin{equation}
\label{eoskt}
\begin{gathered}
K_T \equiv -v \left( \frac{\partial P}{ \partial v} \right)_{T} = \dfrac{v R_{\rm g} T }{\left(v-b_{\rm{p}}\right)^2} \\
+ \dfrac{2}{v^2}\left(-b_{\rm{p}}\left(1-F_{\rm{pr}}\right)\theta \left[R_{\rm g} T\left(1-\exp{ \left( -\frac{N_{\rm{A}} \epsilon_{\rm{p}}}{R_{\rm g} T} \right) }\right)-N_{\rm{A}} \epsilon_{\rm{p}} \right] \left(1-\frac{b_{\rm{p}}}{v}\right)^{\theta-1}-a_{\rm{p}}\right) \\ 
+ \dfrac{b_{\rm{p}}^2}{v^3}\left(1-F_{\rm{pr}}\right)\left(\theta-1\right)\theta \left[ R_{\rm g} T\left(1-\exp { \left(-\frac{N_{\rm{A}} \epsilon_{\rm{p}}}{R_{\rm g} T} \right)}\right)-N_{\rm{A}} \epsilon_{\rm{p}}\right] \left(1-\frac{b_{\rm{p}}}{v}\right)^{\theta-2} .
\end{gathered}
\end{equation} 
\end{widetext}
They chose the parameters $\delta_{\rm p}$ and $\epsilon_{\rm p}$ which provided good matching of the EOS to adsorption isotherm data obtained from GCMC simulations of argon in cylindrical silica nanopores at different pore sizes and temperatures. Using the chosen parameters they calculated the isothermal elastic modulus from Eq.~\ref{eoskt}, these results are shown in Fig.~\ref{fig:eosfigkiso}. Notably, even though the equation is rather simple, having only two fitting parameters, it is able to capture the behavior of the elastic properties seen in simulations across various pore sizes and temperatures \cite{Dobrzanski2020}.

\begin{figure}[H] \centering
\centering
\includegraphics[width=\figwidth]{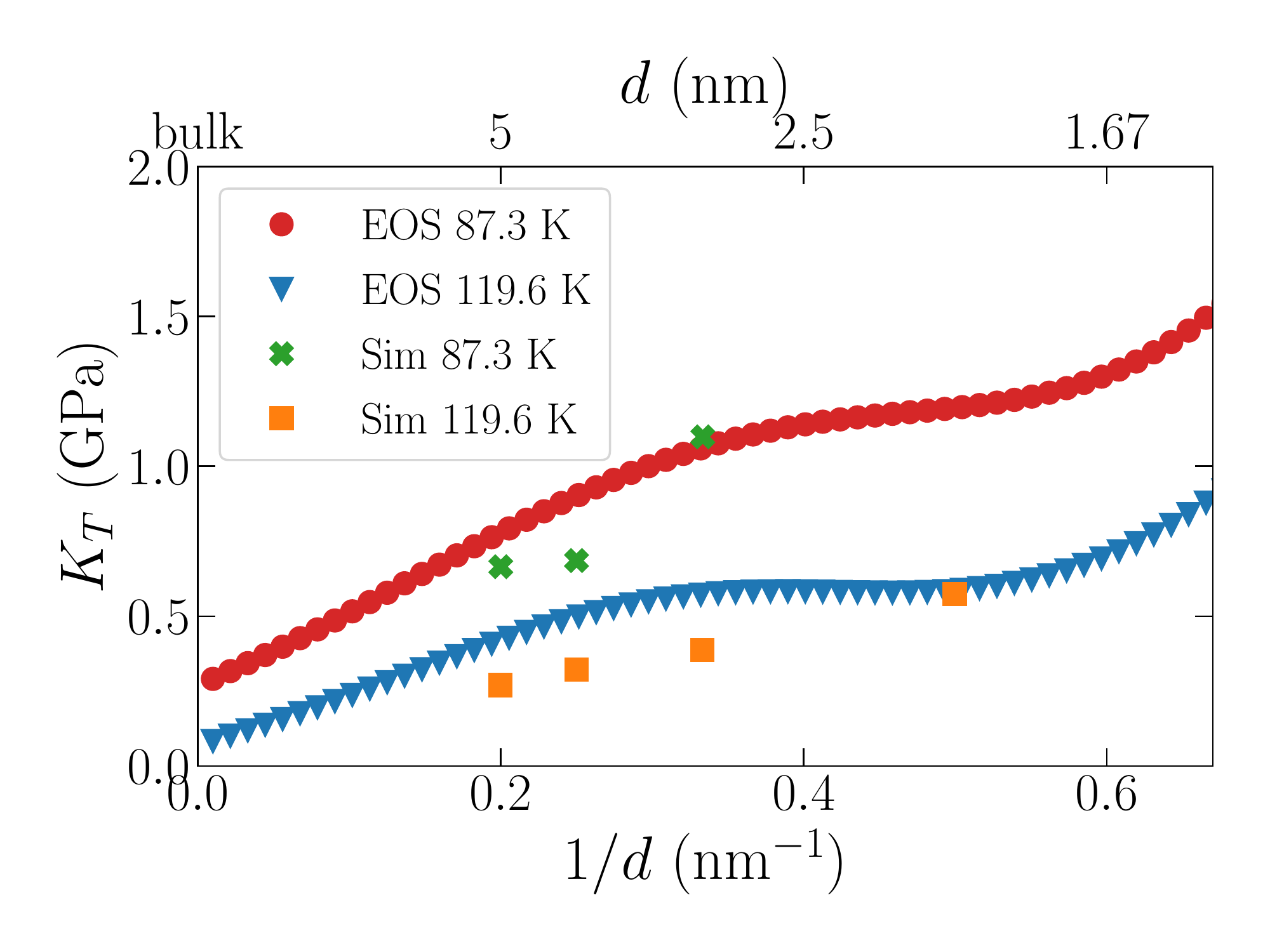}
\caption{Isothermal elastic modulus from GCMC simulation and from EOS at saturation pressure and at temperatures of 87.3 K and 119.6 K plotted versus reciprocal pore size. Given the simplicity of the EOS model, the agreement is very good. Note that at the pore sizes above, ca. $\SI{3}{nm}$, the dependence is linear. Data from Ref.~\onlinecite{Dobrzanski2020}.}
\label{fig:eosfigksat}
\end{figure}

One other relationship that was examined by Dobrzanski et al. was how the elastic modulus depends on the size of the pores. Molecular simulation and DFT works~\cite{Gor2015compr, Dobrzanski2018, Sun2019confinement}, summarized in Sections \ref{sec:MC}, and \ref{sec:Compare}, have shown that the bulk modulus of a subcritical fluid in confinement has a nearly linear relationship with reciprocal pore size, i.e., $K \propto 1/d$. Eq.~\ref{eoskt} predict linear trend for the modulus as a function of $1/d$ for the pore sizes above ca. $\SI{3}{nm}$. For the smaller pore sizes, a slight deviation from linearity is seen (Fig.~\ref{fig:eosfigksat}).

Thus, Dobrzanski et al. were able to show that the trends in adsorption and elastic modulus seen in simulations can be captured using the EOS of Travalloni et al. Showing that an EOS can model the confinement effects on the elastic modulus of the confined fluid was the first step towards a quantitative description of elastic properties, and in turn, wave propagation in fluid-saturated nanoporous media. However, there is still room for improvement, in particular to obtain a quantitative matching of the simulations and EOS across all the temperatures, pressures, and pore sizes for various fluids. The EOS that was used assumes square-well interactions, which has major limitations in replicating behaviors of real fluids. Moreover, it is based on a vdW formalism, which lacks the ability to model temperature dependence on derivative thermodynamic properties. An improved EOS can lead to better modeling of the elastic and other derivative properties across different conditions (i.e., temperature, pressure, and pore size) to be used for practical applications.

\section{Experimental Measurements}
\label{sec:Expt}

The elastic properties of monolithic solid samples can be measured in a relatively straightforward fashion by applying mechanical stresses to a material and measuring the dimensional changes. Clearly, such approaches cannot be applied to confined fluids directly since the measurements would have to be performed on a fluid-saturated nanoporous medium. However, standard static measurements on rocks usually require relatively large strain amplitudes, and thus can be subject to plastic deformations \cite{Jaeger2009, Sinha2001}. Therefore, the elastic properties of fluid-saturated nanoporous media are usually extracted from measuring the speed of elastic waves in the media, typically using ultrasonic frequencies \cite{Gregory1970, Han1986}.

In isotropic solids there are two types of elastic waves. The first is longitudinal waves, which consist of particle motion parallel to the direction of the wave propagation. The longitudinal wave speed, $v_{\rm l}$, is related to the longitudinal modulus $M$. The other type is transverse waves, which consist of particle motion perpendicular to the direction of the wave propagation. The transverse wave speed, $v_{\rm t}$, is related to the shear modulus $G$. The following simple relations describe how these elastic properties along with the material density, $\rho$, determine the wave speeds: 
\begin{equation}
\label{speed}
v_{\rm l} = \left( M/ \rho \right)^{1/2} \qquad \textrm{and}  \qquad  v_{\rm t} = \left( G/ \rho \right)^{1/2}.
\end{equation}
The moduli $M$ and $G$ are related to the bulk modulus $K$:
\begin{equation}
\label{moduli}
K = M - \frac{4}{3}G.
\end{equation} 
Usually, fluids do not support shear stress, therefore $G_{\rm f} =0$, and Eq.~\ref{moduli} indicates there is no difference between the longitudinal modulus $M_{\rm f}$ and bulk modulus $K_{\rm f}$. In this section we use the subscript ``f'' for the fluid properties, subscript ``s'' for the properties of non-porous solid, subscript ``0'' for the properties of dry porous solid, and no subscript for the properties related to the solid-fluid composite (see Figure~\ref{fig:composite}). We do not carry the subscript ``$T$'' (isothermal) used in Section~\ref{sec:Theory}, because the experimentally-measured moduli can be adiabatic as well. 

\subsection{Relating Elastic Properties of Porous Media to the Properties of Confined Phases}
\label{sec:Composite}

\begin{figure}[H] \centering
\includegraphics[width=\figwidth]{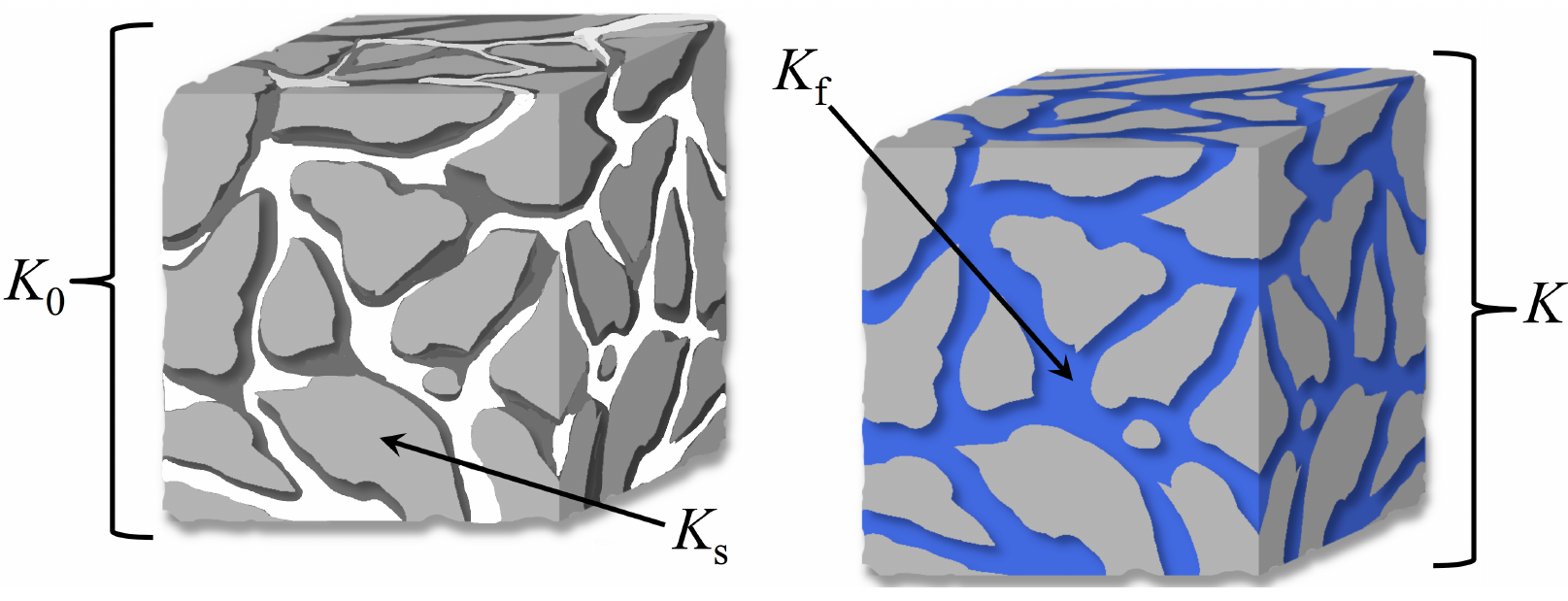}
\caption{Schematic of a porous sample, and bulk moduli of the sample and its constituents: $K_0$ -- modulus of a dry porous sample, $K_{\rm s}$ -- modulus of the non-porous solid (``grains''), $K_{\rm f}$ -- modulus of the fluid, $K$ -- modulus of the fluid-saturated porous sample.}
\label{fig:composite} 
\end{figure}

When the medium of interest is porous and saturated with fluid, the composite properties are determined by those of the solid and fluid constituents. Figure \ref{fig:composite} shows a schematic of a porous medium and denotes the bulk moduli of the constituents involved. In the case of conventional macroporous media, the composite properties are given by the Biot theory of poroelasticity \cite{Biot1956i,Biot1956ii}. When the medium is isotropic and the loads are quasi-static (low-frequency limit), the bulk and shear moduli of the fluid-saturated medium are related to the constituents by Gassmann (or Biot-Gassmann) theory via \cite{Gassmann1951, Berryman1999}
\begin{equation}
\label{no-shear}
G = G_0,
\end{equation}
and
\begin{equation}
\label{Gassmann}
K = K_0 + \frac{\left( 1 - \frac{K_0}{K_{\rm{s}}}\right)^2}{\frac{\phi}{K_{\rm{f}}} + \frac{1-\phi}{K_{\rm{s}}} - \frac{K_0}{K_{\rm{s}}^2}},
\end{equation}
where $G$ is the shear modulus of a fluid-saturated porous sample, $G_0$ is the shear modulus of a dry porous sample, the meanings of various $K$-moduli are indicated in Fig.~\ref{fig:composite}, and $\phi$ is the porosity of the medium. Section \ref{sec:Gassmann} discusses the applicability of the Gassmann theory for ultrasonic experiments on nanoporous glasses.

When the constituent properties ($K_{\rm f}$, $K_{\rm s}$, $K_0$) are known, Eq.~\ref{Gassmann} can predict the properties of the fluid-saturated porous sample. Alternatively, if the modulus of the nonporous solid $K_{\rm s}$ is known,  $K_0$ and $K$ can be measured experimentally, and then Eq.~\ref{Gassmann} can be solved for $K_{\rm f}$. Thus, Eq.~\ref{Gassmann} is the key to relating the experimentally-measurable moduli ($K$, $K_{\rm s}$, $K_0$) to the modulus of the confined fluid $K_{\rm f}$, which cannot be probed in experiments directly. While $K_0$ and $K$ can be measured directly from wave propagation experiments on the dry and saturated samples respectively, the $K_{\rm s}$, corresponding to a nanometer-scale solid pore walls, cannot always be probed in this fashion, and therefore inaccuracy in its value introduces some arbitrariness in the calculation of $K_{\rm f}$.

\subsection{Coupled Adsorption-Ultrasonic Measurements}
\label{sec:Coupled}

\begin{figure}[H] \centering
\centering
\includegraphics[width=\figwidth]{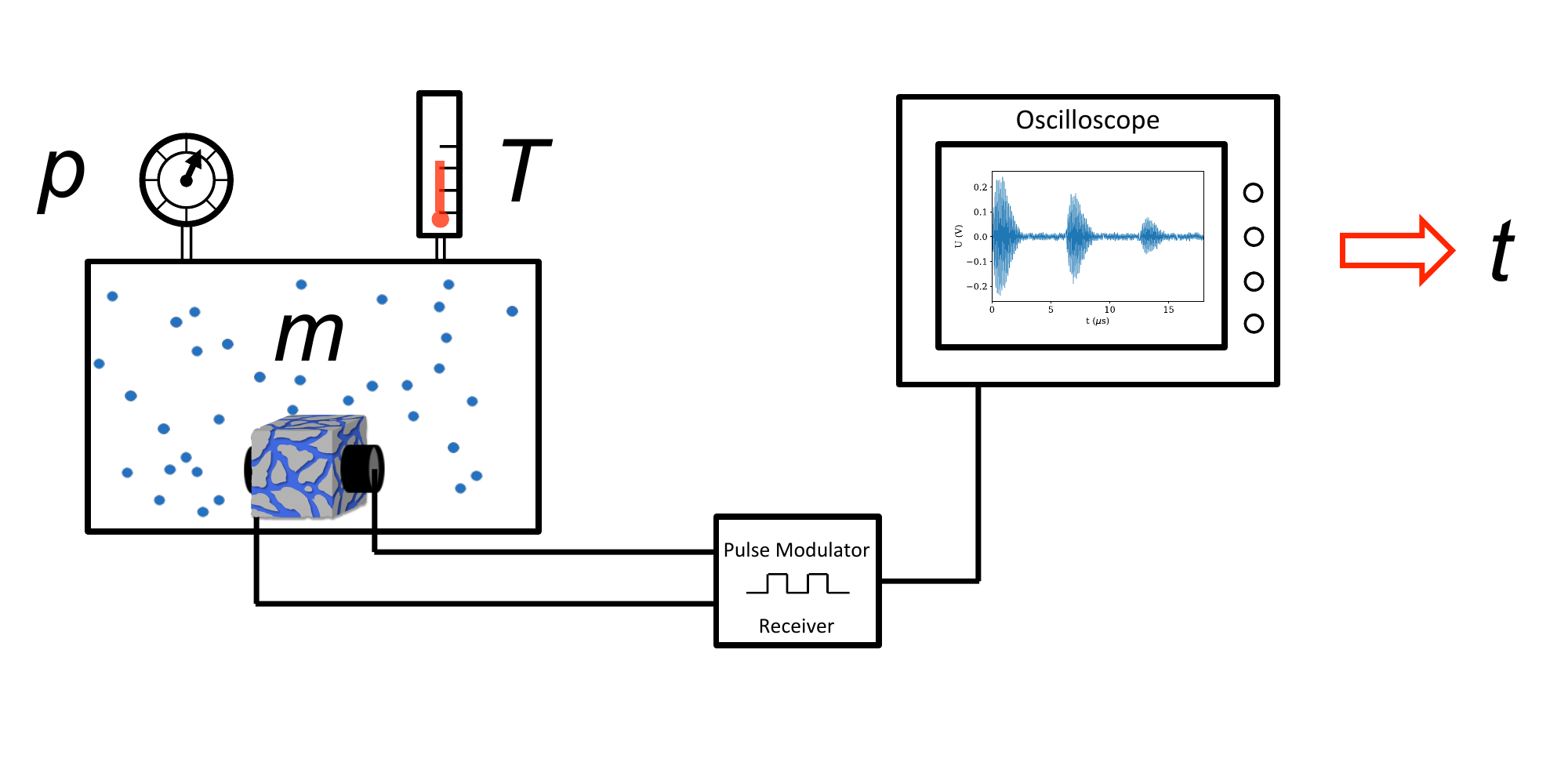}
\caption{Schematic of experimental setup of simultaneous adsorption measurements and ultrasonic wave measurements, such as used by Warner and Beamish \cite{Warner1988}. The temperature $T$ is fixed, and the mass adsorbed is measured as a function of the gas pressure $p$, giving the adsorption isotherm on a nanoporous sample. Ultrasonic transducers (piezo-electric crystals) are bonded to the porous sample and generate the ultrasonic waves. The waves travel through the sample and reflect from the edges of the sample, producing pulse-echo waveforms. The pulse-echo waveforms are displayed on the oscilloscope, where the time between pulse peaks are used to calculate speed of sound.}
\label{fig:ussetup}
\end{figure}

The elastic moduli of a fluid-saturated porous medium (monolithic solid-fluid composite) can be readily derived from the measurements of the sound speed using Eq.~\ref{speed}.  The sound speed is conventionally measured with ultrasonic transducers, a source and a receiver that are attached to the sample surfaces (opposite faces). To secure the uniform filling of the nanoporous medium with the fluid, the samples are gradually filled with condensate by adsorption from the vapor phase, and the speeds of ultrasound propagation through the sample are measured during the adsorption process. A simplified schematic of such an experimental setup is depicted in Figure~\ref{fig:ussetup}.

This design of experiment was proposed by Murphy in 1982, who measured sound speed and attenuation as a function of relative humidity in Massilon sandstone (10 - 100 $\mu$m pores) and compared the results to similar experiments on nanoporous Vycor glass \cite{Murphy1982}. Murphy found that even though the sandstone is 88\% quartz and only 4\% amorphous silica, it had about 6 times greater losses compared to attenuation on Vycor, which is 96\% amorphous silica. Murphy attributed this distinction due to differences in surfaces and pore properties of the materials: Massilon sandstone had flatter pores and rougher surfaces, thus being more compliant and generating more viscous losses compared to Vycor, which has rounder pores with smooth surfaces.

\begin{figure}[H] \centering
\centering
\includegraphics[width=\figwidth]{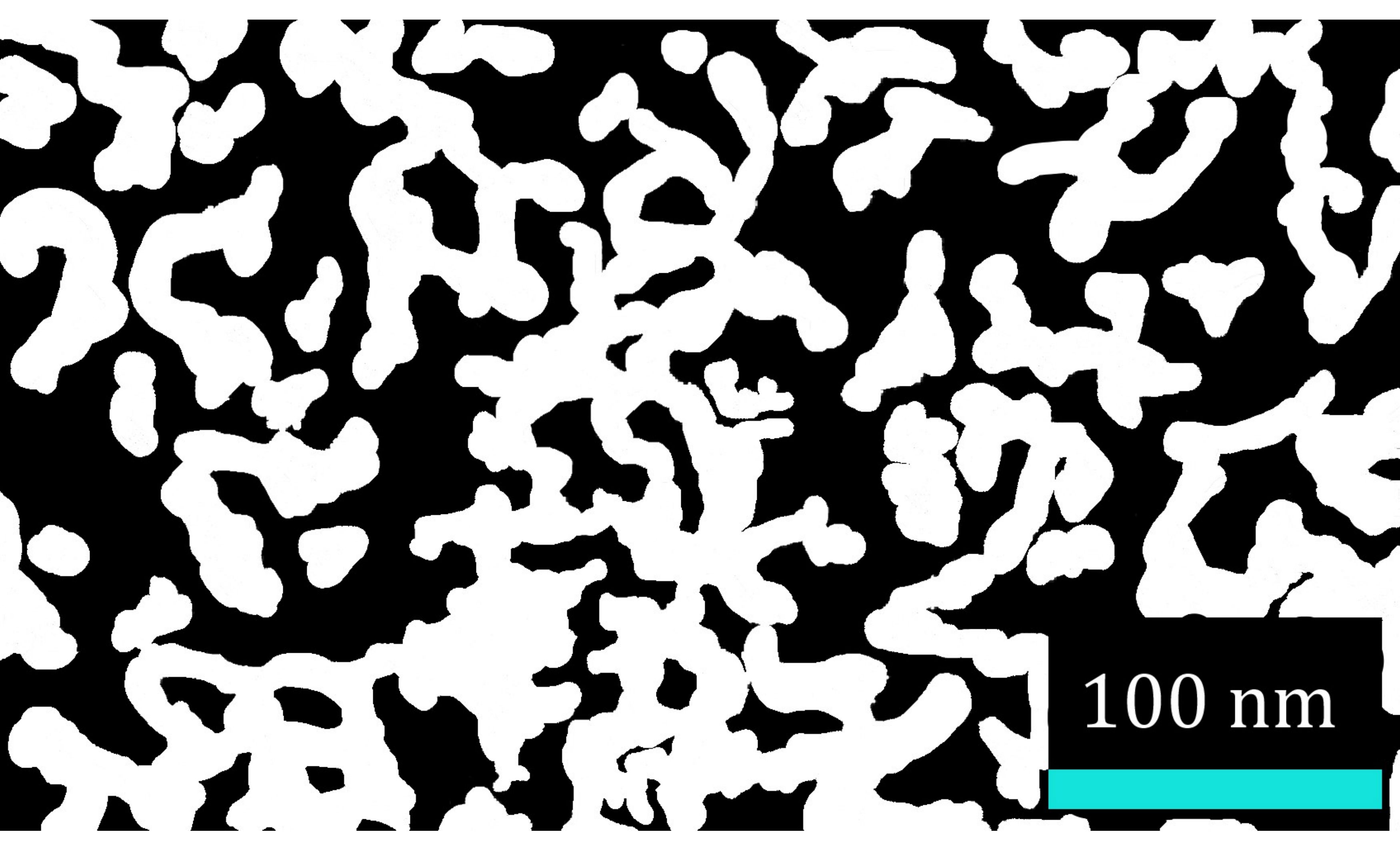}
\caption{Illustrative drawing of a 2-D slice of nanoporous Vycor glass based on the image from Ref.~\onlinecite{Levitz1991}. The white space represents the pore space of the material and the black represents the solid structure.}
\label{fig:Vycor}
\end{figure}

Although Murphy's work was not focused on the confined fluid properties, it has drawn attention towards ultrasonic measurements on Vycor glass samples. Vycor 7930 glass, depicted schematically in Figure \ref{fig:Vycor}, has disordered channel-like pores with a narrow pore size distribution peaked at around $\SI{7}{nm}$ and offers a convenient medium for studying fluids in confinement \cite{Huber2015}. Moreover, unlike many other nanoporous materials, Vycor glass has been manufactured as monoliths. Wherefore, a number of ultrasonic studies of fluids in confinement were performed using Vycor glass as the adsorbent \cite{Warner1986, Warner1988, Molz1993, Page1993, Page1995, Molz1995, Charnaya2001, Borisov2006, Charnaya2008, Borisov2009, Schappert2008}. Finally, Vycor glass is optically transparent, therefore suitable for the comparison of ultrasonic measurements with optical experiments \cite{Page1993, Page1995, Soprunyuk2003, Bonnet2013, Ogawa2013, Ogawa2013hysteretic, Ogawa2015}.

\begin{figure}[H] \centering
\includegraphics[width=\figwidth]{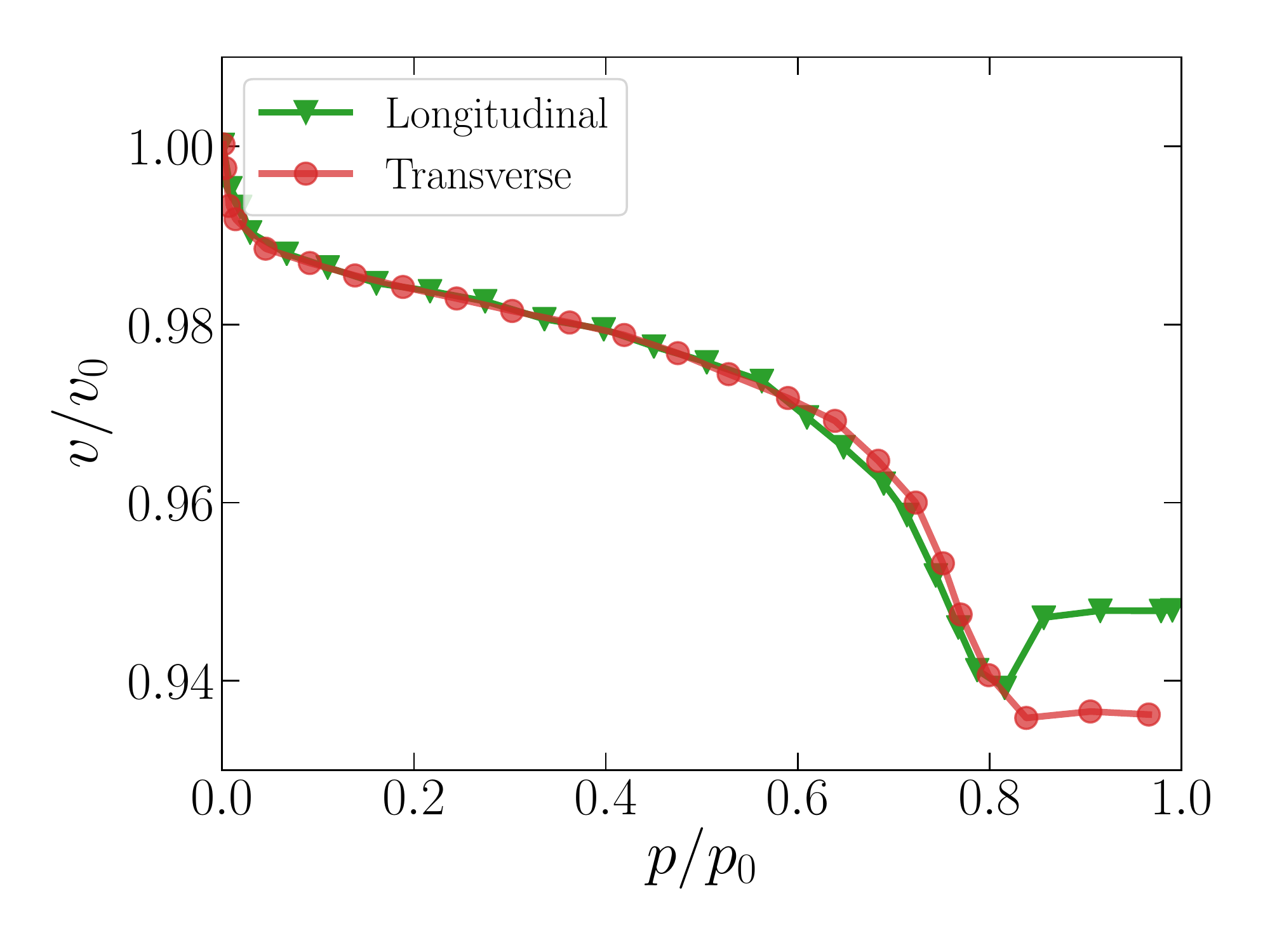} \\
\includegraphics[width=\figwidth]{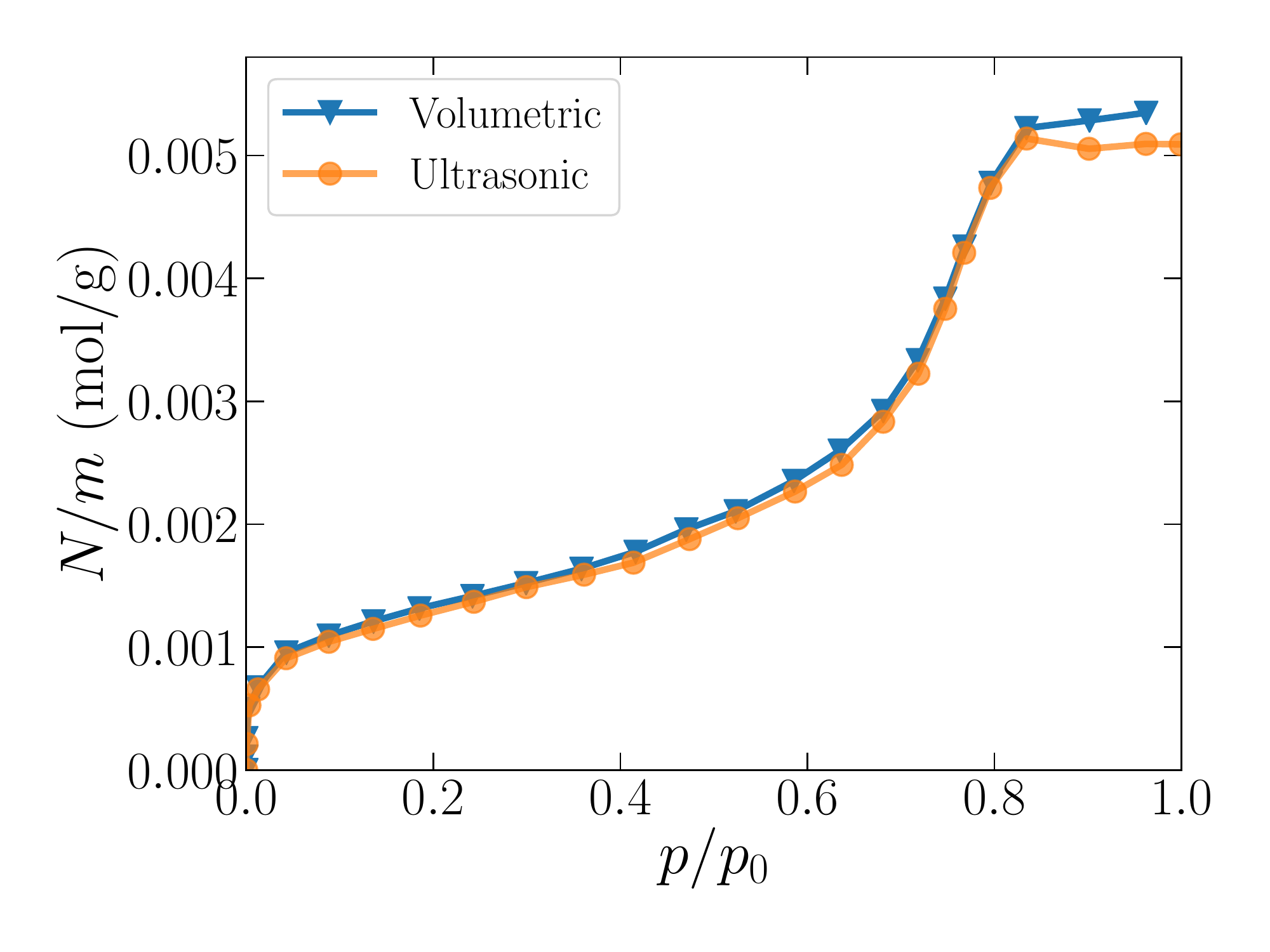}
\caption{Upper panel: speeds of longitudinal and transverse waves through the Vycor sample during nitrogen adsorption. Lower panel: adsorption isotherms measured using conventional volumetric measurements and calculated from the change of the transverse waves speed. Data from Ref.~\onlinecite{Warner1988}.}
\label{fig:Warner}
\end{figure}

A important step was made in 1988 by Warner and Beamish, who used ultrasonic experiments to investigate fluid adsorption on nanoporous samples and their surface area \cite{Warner1988}. Eqs.~\ref{speed} define how the speed of transverse and longitudinal sound waves through a medium depends on its density. When a fluid is allowed to adsorb onto a porous solid, the speed of sound through the medium changes due to the change of its density and, potentially, the change of its elastic modulus. Figure \ref{fig:Warner} (upper panel) shows the speeds of longitudinal and transverse waves through the Vycor glass sample measured by Warner and Beamish as a function of relative vapor pressure. If one assumes, similarly to the bulk fluid, that the shear modulus of the fluid in the pores is zero, the fluid would not contribute to the composite system's shear modulus, then the effective shear modulus of the system would be the same as the shear modulus of the empty porous sample $G = G_0$. This allows straightforward and direct probing of the sample density using ultrasonics via Eqs.~\ref{speed}. Warner and Beamish utilized this concept to relate the amount of fluid adsorbed to the speed of sound, thus proposing an alternative way to measure an adsorption isotherm. Their data, shown in lower panel of Figure~\ref{fig:Warner}, demonstrates that the adsorption isotherms determined from sound speed measurements are fully consistent with adsorption isotherms obtained through volumetric measurements and that the ultrasonic method is also applicable for calculation of the specific surface area. This consistency between the two isotherms justifies the underlying assumption $G = G_0$. It also justifies the assumption that the measurements are not affected by squirt dispersion, as monolithic Vycor samples do not have cracks of aspect ratio $< 0.01$ (See Section~\ref{sec:Gassmann}).

The work by Warner and Beamish~\cite{Warner1988} proposed the use of ultrasonic measurements as an alternative to conventional methods (such as volumetric) for measuring an adsorption isotherm. Moreover, their experimental data, the change of the transit time and the sample mass, can also provide complementary information for the system when both are used together. The resulting change of the longitudinal modulus as a function of the relative vapor pressure can be utilized for calculating the elastic properties of confined fluids. The calculation of the fluid modulus, however, was not reported in their work. It was calculated only recently in Ref.~\onlinecite{Maximov2018} to compare with the predictions of molecular simulation (Section~\ref{sec:Compare}).

\subsection{Probing the Elastic Properties of Confined Fluids}
\label{sec:Probing}

The next important step was the work of Page et al., who combined ultrasonic measurements during vapor adsorption in nanoporous media with optical measurements for hexane adsorption on Vycor glass~\cite{Page1993, Page1995}. The main focus of their work was not on the fluid properties, but on the pore-space, particularly on how the fluid fills the pore-space and how the filled pores are spatially correlated. However, they were the first to analyze the change of the longitudinal modulus of the medium due to the fluid adsorption. Eq.~\ref{speed} gives the following relation between the relative change in transit time $\Delta t/t_0$, the relative change of the sample mass ${\Delta m}/{m_0}$, and the relative change of the longitudinal modulus of the medium $\Delta M/M_0$ (Eq.~4 in Ref.~\onlinecite{Page1995}):
\begin{equation}
\label{PageEq4}
\frac{\Delta M}{M_0} = \frac{\frac{\Delta m}{m_0} - \left[  2\frac{\Delta t}{t_0} + \left( \frac{\Delta t}{t_0} \right)^2 \right]}{\left( 1 + \frac{\Delta t}{t_0} \right)^2},
\end{equation}
where $m_0$ and $\rho_0$ are mass and density of the dry sample respectively.

\begin{figure}[H] \centering
\includegraphics[width=\figwidth]{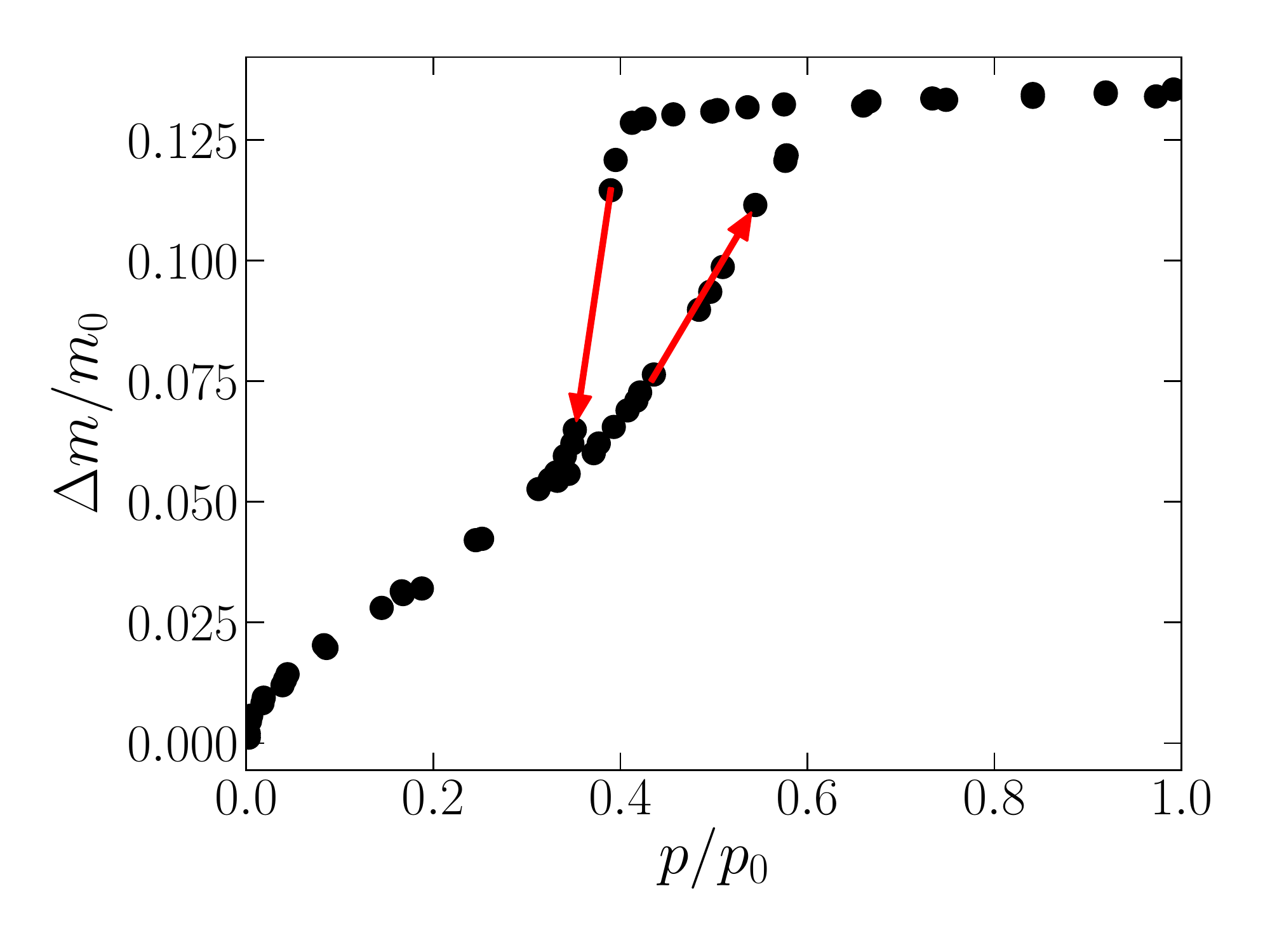}\\
\includegraphics[width=\figwidth]{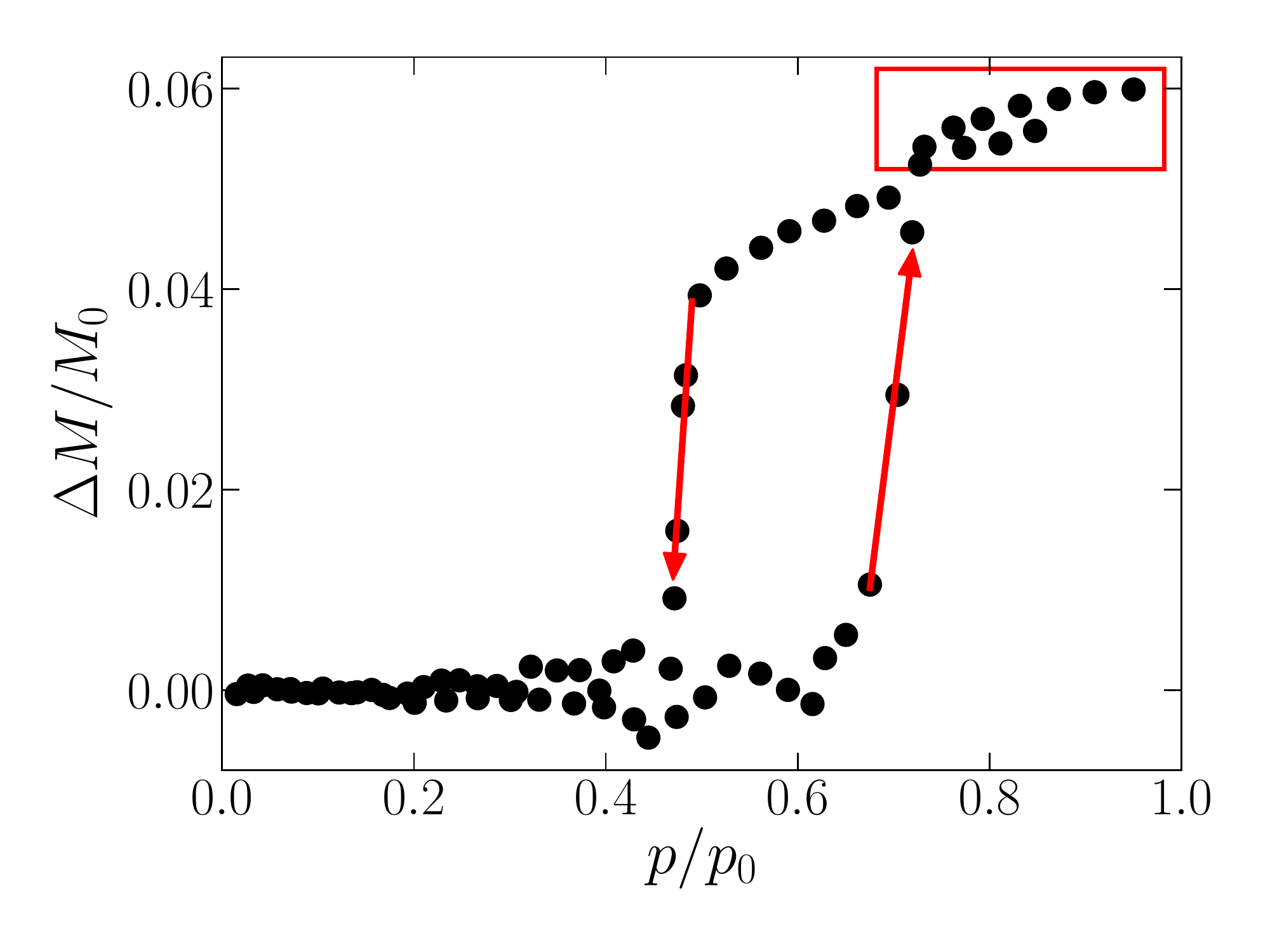}
\caption{Top: adsorption isotherm of n-hexane on a Vycor glass sample. Bottom: relative change of longitudinal modulus of the sample during adsorption. The arrows show the direction of the process -- adsorption and desorption. The rectangle highlights the points after the capillary condensation, when the pores are filled with liquid-like adsorbate. Data from Ref.~\onlinecite{Page1995}.}
\label{fig:Page} 
\end{figure} 

By plotting $\Delta M/M_0$ computed using Eq.~\ref{PageEq4}, Page et al. demonstrated that the longitudinal modulus of a porous sample is approximately unchanged as the vapor pressure increases until the pores are completely filled, at which point there is a rapid increase in the longitudinal modulus. Their data are shown in Figure \ref{fig:Page}: the top panel shows the mass of liquid $\Delta m/m_0$ adsorbed in the porous sample as a function of the relative vapor pressure $p/p_0$, i.e. the adsorption isotherm. The bottom panel shows the associated change in the longitudinal modulus $M$ calculated using Eq.~\ref{PageEq4} from the measured changes of mass density and speed. This plot shows that at relative pressures below $\sim 0.7$, the adsorbed fluid forms a polymolecular film on the pore walls, and the longitudinal modulus of the sample is nearly unchanged. Above $p/p_0 \sim 0.7$, after the pores are filled with liquid by capillary condensation, the modulus increases significantly. Furthermore, after the pores are filled with liquid by capillary condensation, i.e., when the isotherm is practically flat, the modulus $M$ keeps gradually increasing with $p/p_{0}$ and reaches its maximum value at the saturation pressure ($p= p_0$). 

\begin{figure}[H] 
\centering
\includegraphics[width=\figwidth]{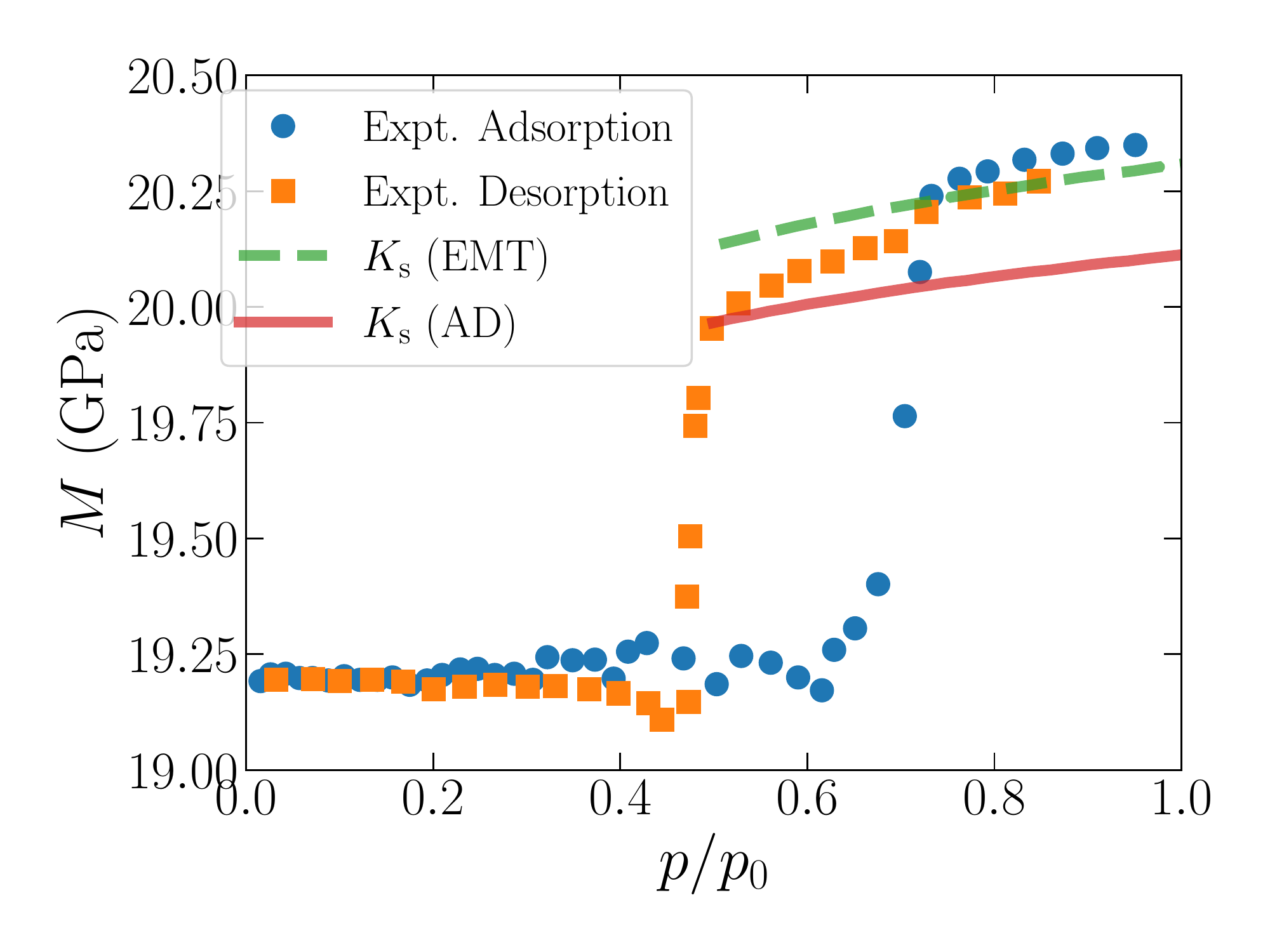}
\caption{Longitudinal modulus $M$ of porous Vycor glass with adsorbed hexane calculated from the ultrasonic data from Page et al.~\cite{Page1995} as a function of relative hexane vapor pressure. The points with circles and squares represent the experimentally obtained modulus during adsorption and desorption respectively. The dashed line represents the calculations based on the $K_{\rm s}$ obtained from adsorption-induced deformation (AD) on quartz and the solid line is using the $K_{\rm s}$ obtained from effective-medium theory (EMT) based on the parameters used by Gor and Gurevich. Data from Ref.~\onlinecite{Gor2018Gassmann}.} 
\label{fig:hexane} 
\end{figure}

To our knowledge, Page et al. \cite{Page1995} were the first to apply the Gassmann equation to analysis of wave propagation in a nanoporous medium. This was done in the calculatation of the elastic modulus of liquid hexane in confinement. For the longitudinal modulus of the sample, the Gassmann equation is conveniently represented as:
\begin{equation}
\label{GassmannM}
M = M_0 + \frac{(K_{\rm s} - K_0)^2 K_{\rm f}}{\phi K_{\rm s}^2 + \left[ (1 - \phi )K_{\rm s} - K_0 \right] K_{\rm f}}.
\end{equation}
Eq.~\ref{GassmannM} provides the value of $K_{\rm f}$ from the data shown in Figure~\ref{fig:Page}. Note that in their calculations, Page et al. used the value of $K_{\rm s}$ for the Vycor sample corresponding to nonporous quartz glass. This difference affected the analysis of the data. Recently, Gor and Gurevich \cite{Gor2018Gassmann} revisited the experimental data from Ref.~\onlinecite{Page1995}, and performed the analysis using the $K_{\rm s}$ value calculated from porosity $\phi$, bulk $K_0$ and shear $G_0$ moduli of the dry sample using the effective medium theory \citep{Kuster1974, Berryman1980} and assuming that the pores are approximately cylindrical in shape. This resulting value of $K_{\rm s}$ was consistent with the earlier work by Scherer \cite{Scherer1986} and much lower than the value for the elastic modulus of quartz glass. As a result, Gor and Gurevich obtained the $K_{\rm f}$ values different from what has been reported in Ref.~\onlinecite{Page1995}, but consistent with the theoretical predictions (we discuss this in detail in Section~\ref{sec:Compare}). The moduli of liquid hexane-saturated Vycor glass sample calculated using the parameters from Page et al.~\cite{Page1995} and from Ref.~\onlinecite{Gor2018Gassmann} are shown in Figure~\ref{fig:hexane}. Irrespective of the value of $K_{\rm s}$ used for calculation of $K_{\rm f}$, a clear trend is seen: the modulus of hexane in the pores is not constant, but changes linearly with the Laplace pressure. This was pointed out in the paper by Page et al.~\cite{Page1995}, and it is in line with the Tait-Murnaghan equation (Eq.~\ref{Tait}), discussed in Section~\ref{sec:Pressure-Modulus}.

Similar studies were reported in a series of papers by Schappert and Pelster~\cite{Schappert2008, Schappert2011, Schappert2013JoP, Schappert2014}. They focused mainly on liquid argon in Vycor glass and obtained the results which are qualitatively similar as in Ref.~\onlinecite{Page1995}. They also related the change of modulus of confined fluid to the adsorption-induced deformation (Section~\ref{sec:Deformation}). It is worth noting that to relate $K$ to $K_{\rm f}$, $K_{\rm s}$, $K_0$, they used an effective medium theory that differs from the Gassmann equation, namely in the following form:
\begin{equation} 
\label{SP-EMT}
K = K_0 + \left( 1 - \frac{K_0}{K_{\rm s }} \right)	K_{\rm{f}},
\end{equation}
where $K_{\rm s}$ for Vycor glass was assumed to be equal to $K_{\rm s}$ for quartz. Their method relies on the assumption that the modulus of the porous sample $K_0$ has a linear dependence on porosity at the low porosity range $\phi \lesssim 0.25$. In a later work,~\cite{Schappert2018} Schappert and Pelster showed that Eq.~\ref{SP-EMT} using the value of $K_{\rm s}$ for quartz gives results close to using Eq.~\ref{Gassmann} with $K_{\rm s}$ calculated as described above (Eq.~\ref{eq_for_K1}) and in Ref.~\onlinecite{Gor2018Gassmann}.

\subsection{Relation between the Ultrasonic Measurements and Adsorption-Induced Deformation}
\label{sec:Deformation}

Adsorption-induced deformation is expansion or contraction of porous materials upon fluid adsorption~\cite{Gor2017review}. Although the magnitude of this deformation is typically small, this phenomenon is ubiquitous. Unless the adsorption is site-specific, the driving force for the deformation is the solvation pressure -- high pressure exerted on pore walls by the confined fluid~\cite{Gor2016Bangham, Gor2016quartz}. The solvation pressure in the pore can be represented as the sum of two contributions: \cite{Gor2010}
\begin{equation}
\label{Psolvation}
P_{\rm s} = P_{\rm sl} + P_{\rm L} ,
\end{equation}
where the first term is related to solid-fluid interactions and the second term is the Laplace pressure:
\begin{equation}
\label{PLaplace}
P_{\rm L } = \frac{R_{\rm g} T}{V_{\rm l}} \log\left( \frac{p}{p_0}\right).
\end{equation} 
Here, $R_{\rm g}$ is the gas constant, $T$ is the absolute temperature, and $V_{\rm l}$ is the molar volume of the liquid phase.  Note that while the first term in Eq.~\ref{Psolvation} is compressive, the second term causes the tensile stresses when the system is in equilibrium with undersaturated vapor at $p < p_0$ (at $p=p_0$ the Laplace pressure term vanishes).

\begin{figure}[H] \centering
\centering
\includegraphics[width=\figwidth]{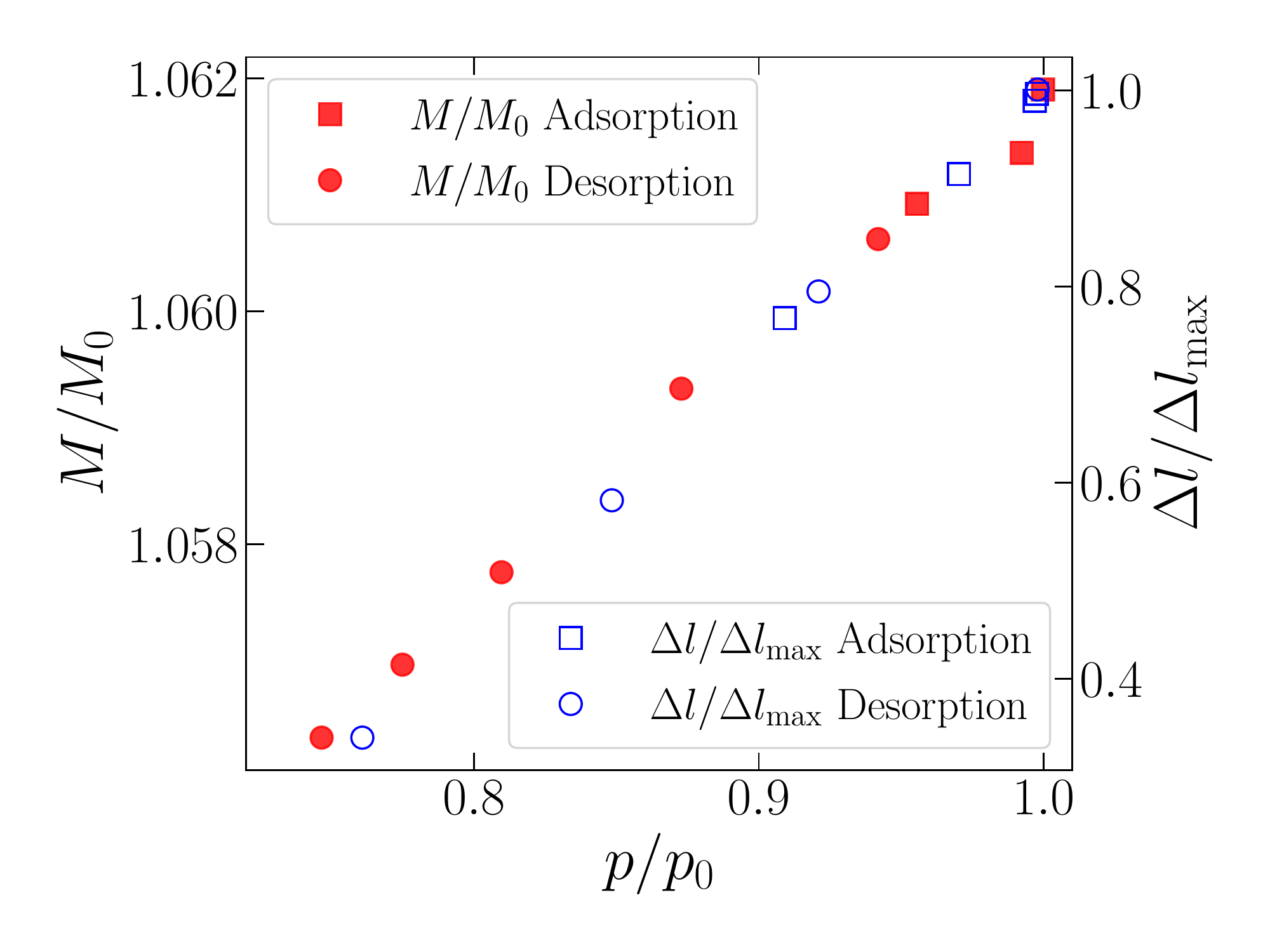}
\caption{Relative change of the longitudinal modulus $M/M_0$ (red filled markers) of Vycor glass and relative elongation of the sample $\Delta l/ \Delta l_{\rm max}$ (blue open markers) as a function of argon vapor pressure. With a proper choice of scales on the $y$-axes, the data collapse into a single curve, suggesting a linear relation between the $M/M_0$ and  $\Delta l/ \Delta l_{\rm max}$. Data from Ref.~\onlinecite{Schappert2014}.}
\label{fig:Schappert2014}
\end{figure}

Recent experiments by Schappert and Pelster showed a correlation between adsorption-induced deformation and the change of the elastic modulus of the fluid-saturated sample. They measured the speed of ultrasound propagation in a porous glass sample in the course of argon adsorption~\cite{Schappert2008, Schappert2011, Schappert2013JoP}. From the ultrasonic measurement, they calculated the relative change of the longitudinal modulus of the sample, shown with red filled markers in Figure \ref{fig:Schappert2014}. Furthermore, they complemented the ultrasonic measurements by measurement of adsorption-induced deformation~\cite{Schappert2014} -- relative elongation of the sample as a result of fluids adsorption, which is depicted by the open markers in Figure \ref{fig:Schappert2014}. Displayed on the same plot, these points demonstrate a linear relation between the change of the fluid modulus and the elongation of the sample. The linear relation between the change in modulus and deformation confirms the linear relation between the modulus and the Laplace pressure, which was earlier observed by Page et al.~\cite{Page1995}, consistent with the Tait-Murnaghan equation (Eq.~\ref{Tait}). 

Since adsorption-induced strains of mesoporous materials at high relative pressures have a logarithmic dependence on the relative pressure, the experiments on adsorption-induced deformation provide a straightforward way to estimate the elastic properties of solid samples, in particular, the solid modulus $K_{\rm s}$, which is necessary for application of Gassmann equation for the analysis of ultrasonic data measured on a fluid-saturated sample. This approach was used by Gor and Gurevich~\cite{Gor2018Gassmann} to analyze the experimental data from Refs.~\onlinecite{Page1995, Schappert2014}.

\begin{figure}[H] \centering
\centering
\includegraphics[width=\figwidth]{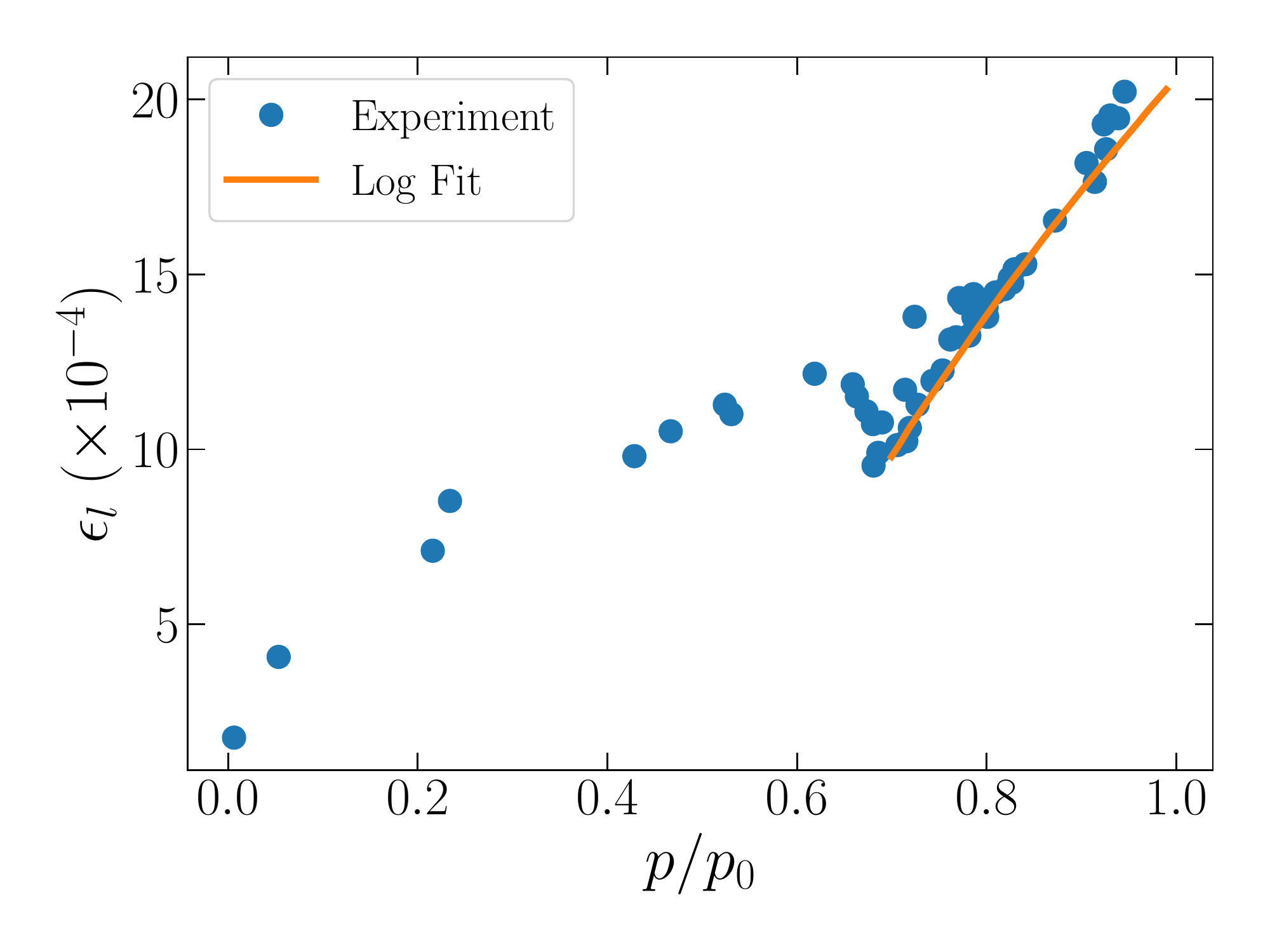}
\caption{Experimental data on adsorption-induced deformation from Ref.~\onlinecite{Amberg1952} -- water adsorption on a Vycor glass sample at $\SI{291.9}{K}$, used in Ref.~\onlinecite{Gor2018Gassmann} for calculation of solid elastic modulus $K_{\rm s}$. The circles represent the experimental linear strain. The solid line represents a log fit to the data at higher vapor pressure when the pores are filled with water. Data from Ref.~\onlinecite{Gor2018Gassmann}.}
\label{fig:Amberg}
\end{figure}

For a sample saturated with a fluid at a constant temperature, the term $P_{\rm sl}$ is constant, so Eq.~\ref{PLaplace} gives a logarithmic dependence of linear strain of the porous sample $\epsilon_l$ with respect to $p/p_0$, as shown in Figure \ref{fig:Amberg} and observed for all mesoporous materials \citep{Gor2017review}. This dependence is often described using a special elastic modulus related to this process, the so-called ``pore-load modulus'' $\mathcal{M}_{\rm PL}$ \citep{Prass2009,Gor2015modulus} as a proportionality constant in the linear relation between the solvation pressure $P_{\rm f }$ and measured $\epsilon_l$. $\mathcal{M}_{\rm PL}$ can be related to elastic moduli using the following equation \cite{Mackenzie1950}:
\begin{equation}
\label{eq_for_K1}
\frac{3}{\mathcal{M}_{\rm PL}}+\frac{1}{K_{\rm s}}=\frac{1}{K_0}. 
\end{equation}
When $\mathcal{M}_{\rm PL}$ and $K_0$ are known from experimental measurements, Eq.~\ref{eq_for_K1} can be used to estimate $K_{\rm s}$. Note that the value of $K_{\rm s}$ for Vycor glass calculated from the adsorption-induced deformation data from Ref.~\onlinecite{Amberg1952}, agreed well with the calculation based on the values of $K_0$, $G_0$, and $\phi$ using the effective medium theory \cite{Gor2018Gassmann}.

High pressure in the confined fluid is exerted on the solid, therefore, according to the Tait-Murnaghan Eq.~\ref{Tait}, similarly to the change of the bulk modulus of the fluid, there could be a change of the bulk modulus of the solid. Ref.~\onlinecite{Gor2018Gassmann} estimated this effect for quartz, based on the constant $\alpha$ from Ref.~\onlinecite{Anderson1966}. Because $\alpha$ for solids is noticeably smaller than for fluids, the effects of pressure on the solid could be neglected. The negligible change of the shear modulus of the nanoporous sample when it is filled with fluid also suggests that the high pressure in the pores does not appreciably affect the elastic constants of the solid constituent.

\begin{figure}[H] \centering
\centering
\includegraphics[width=\figwidth]{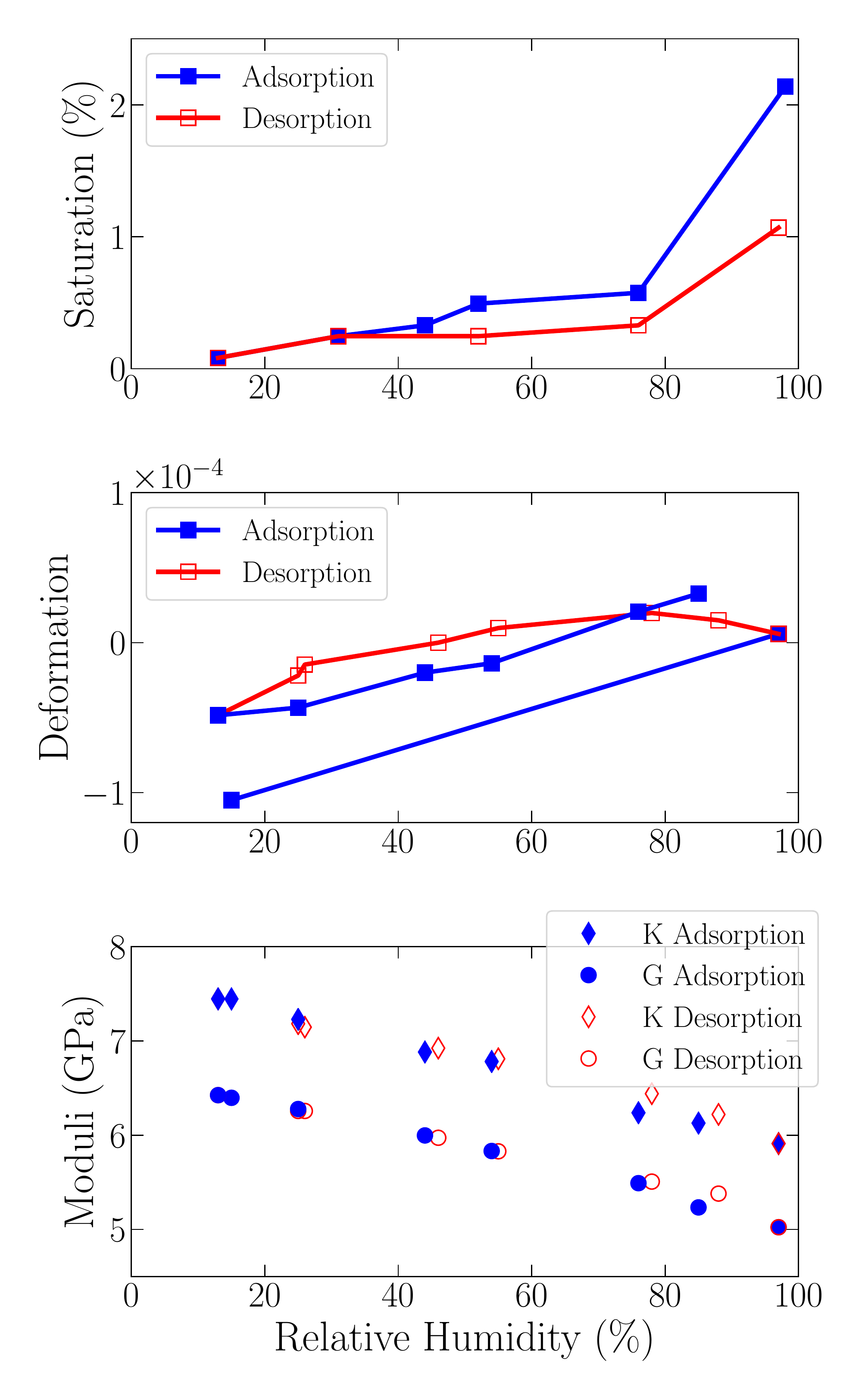}
\caption{Top to bottom: Saturation, deformation, and elastic moduli of a Bentheim sandstone sample during water adsorption (solid markers) and desorption (empty markers) as a function of relative humidity. Data from Ref.~\onlinecite{Yurikov2018}.}
\label{fig:Yurikov}
\end{figure}

Another correlation between deformation and ultrasound propagation has been reported for water adsorption on sandstones. A number of studies have reported a significant reduction of ultrasonic speeds, and/or increase of ultrasonic attenuation in vacuum-dry sandstones, upon imbibition of very small amounts of water \cite{Wyllie1962, Hardin1963, Pandit1979, Tittmann1980, Clark1980, Murphy1982PhD, Knight1992, Tutuncu1993, Pimienta2014, Hossain2019, Tiennot2020, Tadavani2020}. This effect is not entirely understood, but is commonly attributed to the adsorption of water at very thin (likely nano-scale) contacts between adjacent grains. Water adsorption creates solvation pressure, which pushes the adjacent grains away from each other, thus reducing the contact stiffness, which in turn reduces the elastic moduli and increases ultrasonic attenuation. Recently, this mechanism was corroborated by Yurikov et al.~\cite{Yurikov2018}, who showed that the reduction of the moduli caused by water imbibition is accompanied by an expansion of the sample size broadly consistent with the expected deformation caused by solvation pressure. Figure \ref{fig:Yurikov} shows saturation of a sample with water, measured deformation, and elastic moduli as functions of the relative humidity.

\subsection{Freezing in the Nanopores and Shear Modulus of Confined Phases}
\label{sec:Shear}

\begin{figure}[H] \centering
\centering
\includegraphics[width=\figwidth]{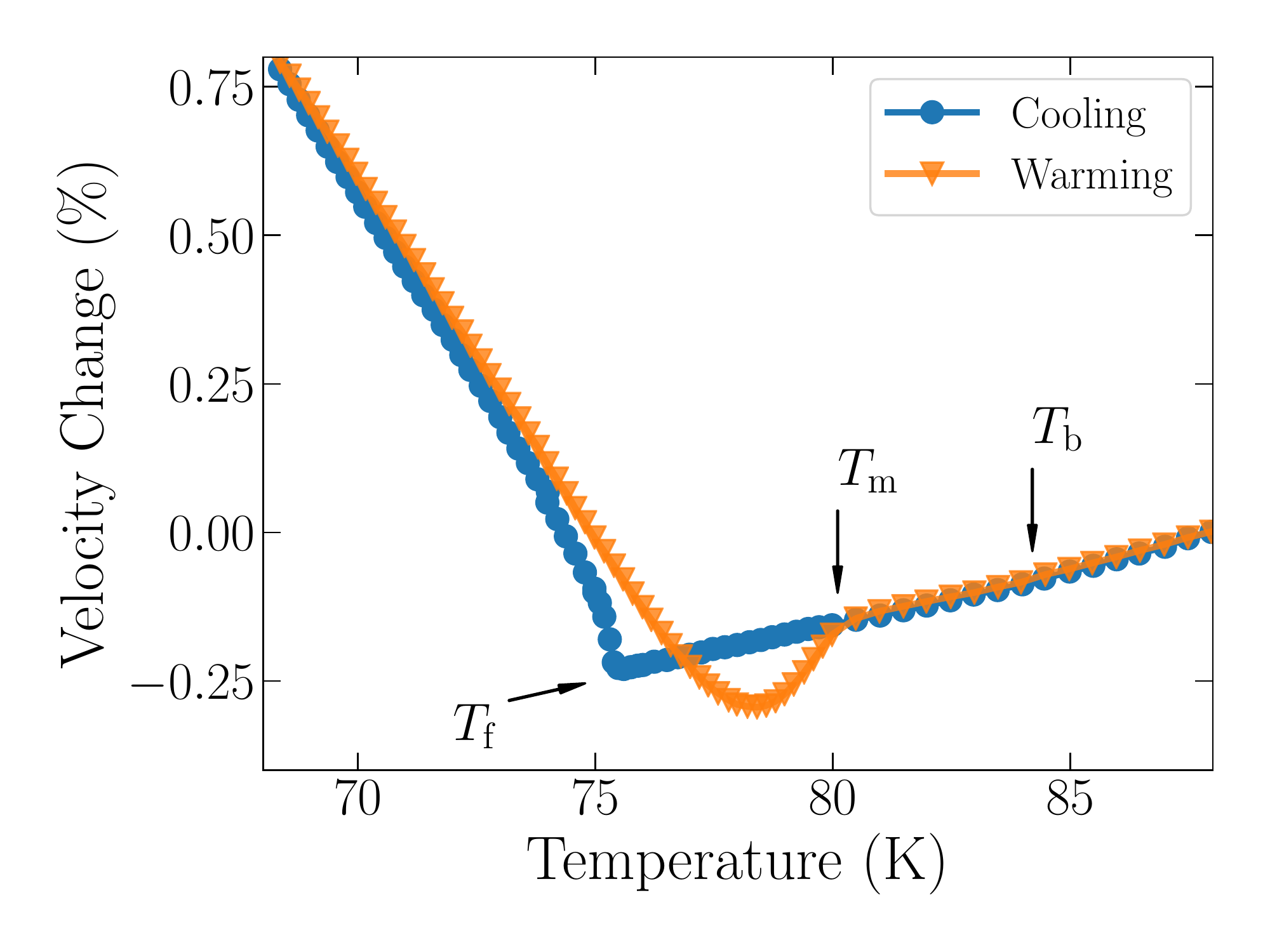}
\caption{Relative change of the sound speed of argon-saturated Vycor sample showing a pronounced hysteresis between freezing and melting. $T_{\rm b}$ is the bulk melting temperature, $T_{\rm f}$ is the onset of freezing upon cooling, and  $T_{\rm m}$ is the completion of melting upon heating. Data from Ref.~\onlinecite{Molz1993}.}
\label{fig:Molz}
\end{figure}

Bulk solid phases are typically stiffer than the same substances in fluid phases: any matter in solid form has a finite shear modulus, hence the longitudinal modulus of solid is higher than in liquid state (see Eq.~\ref{moduli}). In addition, the bulk modulus of a matter in solid state is often higher too. It also applies to confined phases: when a fluid freezes in the pores, its elastic properties noticeably change. This phenomena is seen clearly in speed of wave propagation measurements. This signature of phase transitions has been used in a number of works to monitor the freezing of fluids in confinement, such as helium \cite{Beamish1983, Beamish1984}, argon \cite{Molz1993, Schappert2008, Schappert2011, Schappert2013JoP, Schappert2013PRL}, nitrogen \cite{Schappert2013N2}, oxygen \cite{Schappert2016JPCC}, mercury \cite{Charnaya2001, Borisov2006}, and alkanes \cite{Borisov2009, Schappert2015EPL}. However, many of these works did not quantify the elastic properties of confined phases. Instead, their focus was on the change of the sound wave speed or of the composite modulus.  

Molz et al. \cite{Molz1993} utilized the data on the transverse ultrasonic waves and demonstrated that the sound speed changes gradually in a broad temperature range (broader than the peak on the calorimetric measurements). Their data is shown in Figure \ref{fig:Molz}. During cooling starting from \SI{88}{K}, the speed is gradually decreasing as a result of thermal contraction of liquid argon. At the bulk freezing point $T_{\rm b} = \SI{84}{K}$, there are no appreciable changes in the signal, but at the temperature $T_{\rm f} = \SI{75.55}{K}$  there is a sudden increase in the speed which indicates the onset of the freezing. Note that their measurements suggest that the shear modulus for liquid argon was zero.

\begin{figure}[H] \centering
\centering
\includegraphics[width=\figwidth]{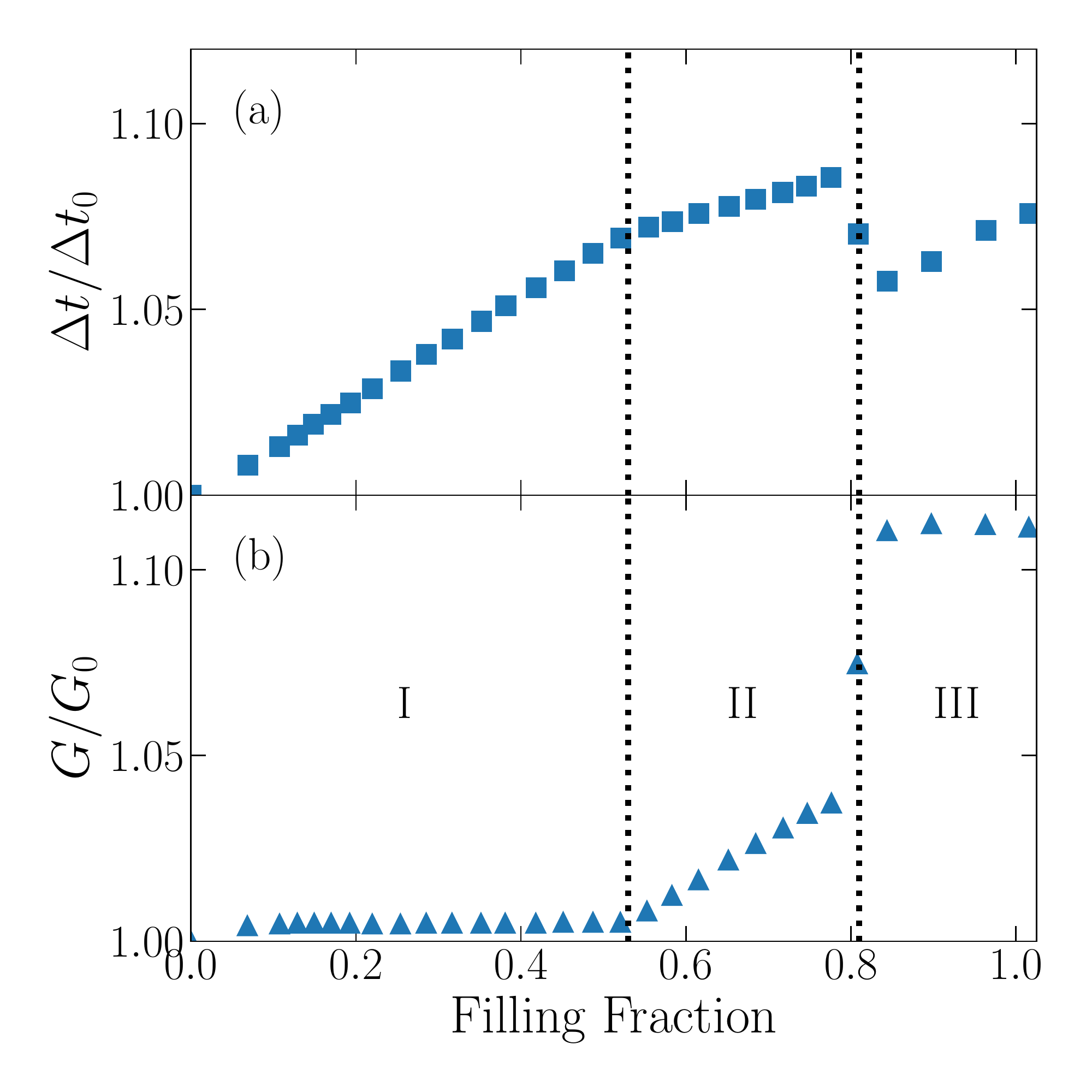}
\caption{(a) Ultrasonic transit time (relative to the transit time of the unfilled sample), and (b) ratio of effective shear modulus $G$ to the shear modulus of the empty sample $G_0$ on adsorption of argon at $T = \SI{72}{K}$. The process of freezing starts above a filling fraction of 0.53. Data from Ref.~\onlinecite{Schappert2008}.}
\label{fig:Schappert2008}
\end{figure}

The ultrasonic study of freezing of liquid argon in confinement was revisited by Schappert and Pelster \cite{Schappert2008}. They determine that there are three regions of filling fraction which have differing behavior for argon below its bulk freezing point. In the first region, using the ultrasonic measurements they found that the shear modulus of the Vycor sample with adsorbed argon does not change when there are fewer than about 3 to 4 adsorbed layers of argon. When the pore is filled past this region of filling, there is a linear increase in shear modulus in the second region (II in Figure~\ref{fig:Schappert2008}). In the third region, when the pores become completely filled, the shear modulus increases abruptly and then remains constant upon further increase of the filling fraction.

Recently, the experimental data from Ref.~\onlinecite{Schappert2014} were revisited by Sun et al., who explored the applicability of elastic effective medium theories, which are routinely used for macroporous media, for the analysis on nanoporous Vycor glass filled with liquid and solid argon~\cite{Sun2019}. In particular, Sun et al. \cite{Sun2019} showed that at $\SI{74}{K}$, under an assumption of spheroidal pore geometry, predictions of the differential effective medium (DEM) theory \cite{Cleary1980, Norris1985, Zimmerman1991} show reasonable agreement with the measured shear modulus of Vycor filled with solid argon, but underestimate its bulk modulus. Moreover, the measured bulk modulus of the Vycor filled with solid argon at $\SI{74}{K}$ is close to the bulk modulus of Vycor filled with liquid argon at $\SI{80}{K}$, despite the fact that bulk modulus of the bulk solid argon is  approximately 1.8 times higher than that for bulk liquid argon~\cite{Keeler1970}. This suggests that the bulk modulus of the confined solid argon at $\SI{74}{K}$ (which is near the melting point of confined argon of $\SI{76}{K}$) may be close to the bulk modulus of liquid argon, and hence significantly lower than for bulk solid argon.

\begin{figure}[H] \centering
\centering
\includegraphics[width=\figwidth]{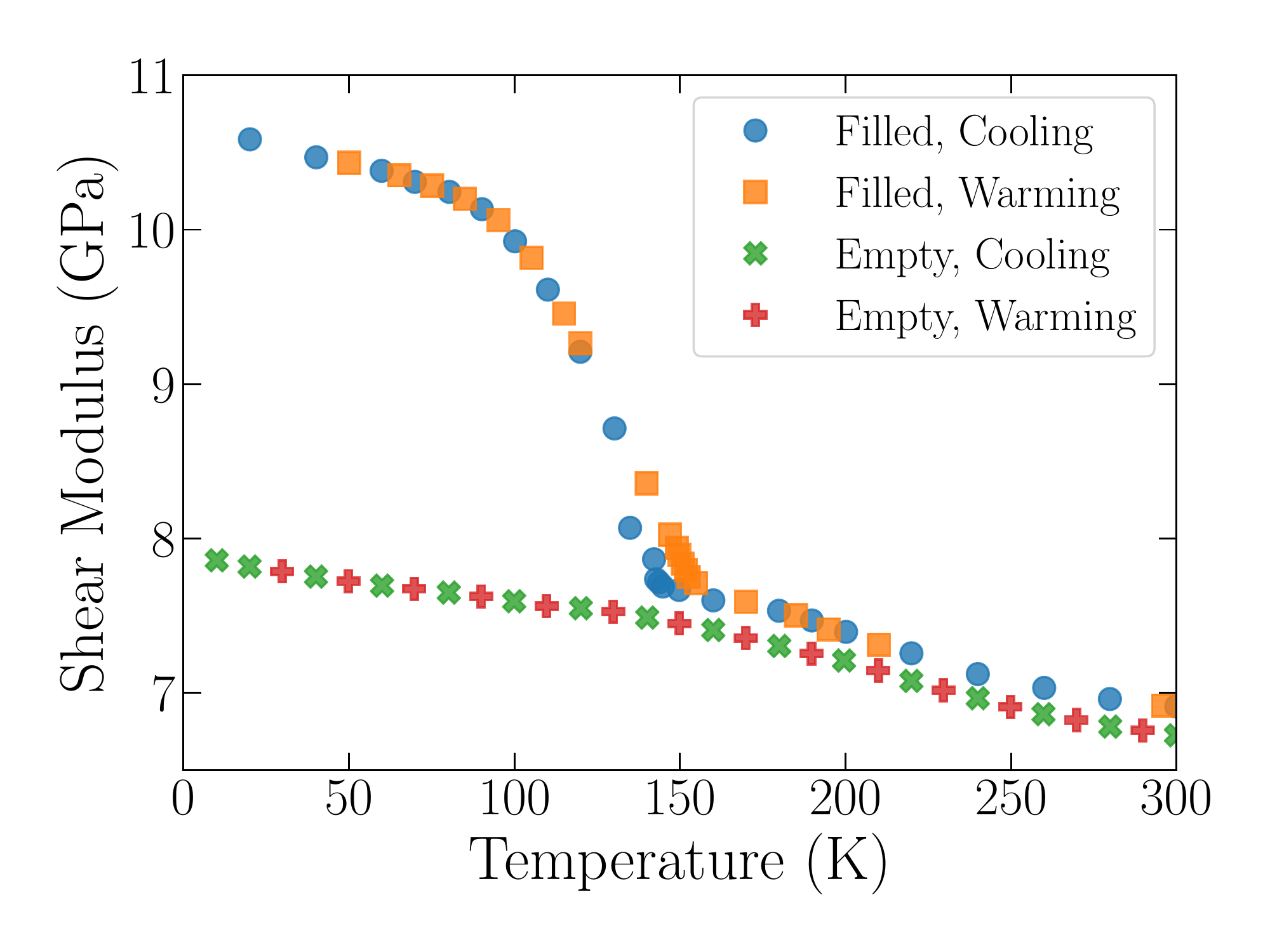}
\caption{Shear modulus of a Vycor glass sample saturated with hexane as a function of temperature. Even at $T > \SI{160}{K}$, when hexane is in liquid phase, the shear modulus exceeds the value for the dry sample. It suggests that unlike argon and nitrogen, confined liquid heptane has a non-zero shear modulus. Data from Ref.~\onlinecite{Schappert2015EPL}.}
\label{fig:Schappert2015}
\end{figure}

Schappert et al. used transverse waves to probe confined fluids which have more complex structure, n-heptane, and n-nonane \cite{Schappert2015EPL}. Figure \ref{fig:Schappert2015} shows one of the results from their work: the shear modulus of the solid sample saturated with heptane exceeds the shear modulus of the dry sample even at temperatures above the confined melting point. Similar observations were made for nonane \cite{Schappert2015EPL}. It suggests that liquid heptane and nonane, when confined in the pores of Vycor glass, have non-zero shear moduli. This conclusion differs from the expectation for bulk liquids and from observations of confined liquid nitrogen \cite{Warner1988} and argon \cite{Schappert2014}. At the same time, this is consistent with the classical surface force measurements for the fluids confined between two parallel planes: when the gap between the planes is on the order of a nanometer, such measurements show that the fluid has the shear viscosity exceeding the bulk value by seven orders of magnitude \cite{Granick1991}. Such dramatic increase of the shear forces, could have an effect on the shear modulus measured in ultrasonic experiments at frequencies of $\SI{7}{MHz}$ \cite{Schappert2015EPL}.

\subsection{Applicability of Gassmann Theory for Nanoporous Media}
\label{sec:Gassmann}

Since Gassmann theory is key to relating the ultrasonic measurements on porous samples to the properties of the confined fluid \cite{Gor2018Gassmann}, it is worth discussing the applicability of this theory for the types of systems such as the experiments on liquid nitrogen confined in Vycor glass from Ref.~\onlinecite{Warner1988}, as an example case study (Section~\ref{sec:Coupled}). First, the frequencies in the range of 1-10 MHz are low enough to neglect the wave scattering on nanopores. Indeed, the characteristic wavelength can be estimated as $\lambda =\frac{v}{(2 \pi f)} \simeq \frac{\SI{851}{m/s}}{\SI{2 \pi E7}{Hz}} = \SI{1.4E-5}{m}$, using the speed corresponding to the bulk liquid nitrogen at normal boiling temperature \cite{Span2000}. Even in this case the wavelength $\lambda$ exceeds the characteristic pore size $\SI{1}{nm}- \SI{10}{nm}$ by 3-4 orders of magnitude. Thus, the wave propagation is ballistic -- it does not scatter and probes fluid-saturated nanoporous medium as a uniform medium \cite{Page1995diffusion}. This distinguishes ultrasonic experiments from experimental techniques based on X-ray or neutron scattering, which have wavelengths comparable to the molecular dimensions and are widely used for probing confined fluids at the molecular level \cite{Melnichenko2015, Morineau2020chapter}. Although these methods have not been applied for probing the elastic properties in molecular fluids, X-ray scattering has been recently utilized for probing the local compressibility of confined colloidal fluid \cite{Nygaard2016, Nygaard2016Opinion} (See Section~\ref{sec:Compare}).

Second, an important restriction of Gassmann (as well as Biot) theory is that the fluid pressure is uniform within the pore space. This requires that shear stresses in the fluid be negligible, that is, the signal frequency is lower than the  crossover frequency of the so-called squirt dispersion  $f_{\rm sq}$, which is on the order $\alpha_{\rm r}^3 G_0/(2 \pi \eta)$, where $\alpha_{\rm r}$ is the typical aspect ratio of the pores or cracks and $\eta$ is the fluid viscosity. For spherical or cylindrical pores, $\alpha_{\rm r} = 1$ and hence $f_{\rm sq}=\SI{10}{THz}$, but $f_{\rm sq}$ can be many orders of magnitude smaller if the solid sample contains thin cracks with $\alpha_{\rm r}$ on the order 0.001 ~\cite{Mavko1975, Mavko1991, Gurevich2010, Muller2010}. If such cracks are present, the shear modulus of the fluid-saturated medium deviates from that in the dry medium and depends on the fluid bulk modulus \cite{Mavko1991, Gurevich2009, Gurevich2010}. However, measurements on Vycor glass (often used in combined adsorption-ultrasonic experiments) saturated with liquid nitrogen or argon show no effect of capillary condensation on the shear modulus \cite{Warner1988, Schappert2014}, suggesting that Vycor contains no such cracks.

The third condition is related to the applicability of the low-frequency limit of Biot theory~\cite{Biot1956i}. The characteristic frequency with respect to which the experimental frequency can be considered low (Gassmann limit), can be estimated as~\cite{Biot1956i} $f_{\max} = \frac {\eta }{\pi \rho_{\rm f}\delta_{\max}^{2}} \simeq \SI{1}{GHz}$, where $\delta_{\rm max} \approx 7-8$~nm is the viscous skin depth considered as the maximum pore diameter for the Vycor sample, $\rho _{\rm{f}} = \SI{807}{\kg\per\cubic\meter}$ is the fluid density~\cite{Span2000}, $\eta = \SI{163}{\micro\Pa\second}$ is the dynamic viscosity for nitrogen in bulk at temperature $T=\SI{77}{K}$ and pressure $P=\SI{0.1}{MPa}$~\cite{Lemmon2004}. Therefore, the frequencies of ca. 10~MHz, used in Ref.~\onlinecite{Warner1988} for transverse and longitudinal waves, can be considered low. Thus, more generally, when the pore sizes are in the nanometer range, the frequencies of up to tens of MHz can be typically considered low and fall under the limit of Gassmann theory. 

\section{Relating Experiment and Theory}
\label{sec:Compare}

While a number of papers reported theoretical findings on elastic properties of confined fluids \cite{Bratko2001, Coasne2009, Martini2010, Strekalova2011, Strekalova2012, Rickman2012, Sun2014, Evans2015, Evans2015PRL, Keshavarzi2016, 
Evans2017}, and another number of papers reported experimental measurements of elastic properties of fluid-saturated nanoporous solids \cite{Murphy1982, Beamish1983, Beamish1984, Warner1986, Warner1988, Molz1993, Page1993, Page1995, Molz1995, Charnaya2001, Borisov2006, Charnaya2008, Borisov2009}, up until recently the connection between theory and experiment has not been made. A series of publications by Schappert and Pelster reported ultrasonic experiments on Vycor glass saturated with liquid argon \cite{Schappert2008, Schappert2011, Schappert2013PRL, Schappert2013JoP, Schappert2014, Schappert2014Langmuir, Schappert2016, Schappert2016correlation, Schappert2018, Schappert2018liquid}, which is an excellent system for molecular modeling because interactions of argon atoms with each other and with glass surfaces can be readily modeled by simple Lennard-Jones potentials. Their work stimulated Gor and co-workers to focus on DFT and MC simulations for this system \cite{Gor2014, Gor2015compr, Gor2016Tait, Gor2017Biot, Dobrzanski2018, Dobrzanski2020}, and to make a step towards the comparison of simulations to ultrasonic data \cite{Gor2018Gassmann, Maximov2018}. 

Ref.~\onlinecite{Gor2014} presented the calculation of isothermal modulus of confined liquid argon based on theoretical adsorption isotherms predicted by QSDFT (see Section \ref{sec:Isotherm} for details). The key result was the logarithmic relation between the modulus and the vapor pressure of the adsorbing argon, shown in Fig.~\ref{fig:Gor2014}, close to that which was measured by Schappert and Pelster \cite{Schappert2014}. The agreement remained qualitative for the following two reasons. The first one is related to the inconsistency of the effective medium theory (Eq.~\ref{SP-EMT}) from Ref.~\onlinecite{Schappert2014} with the widely accepted Gassmann theory~\cite{Gor2018Gassmann}. The second one is due to the systematic error in compressibility predictions of QSDFT. Unlike the calculations based on Monte Carlo simulations, QSDFT for liquid argon did not predict the correct bulk liquid compressibility in the limit of large pores.~\cite{Gor2017Biot}. Nevertheless, Ref.~\onlinecite{Gor2014} was the first work that demonstrated the relation between the compressibility of a confined fluid predicted by molecular modeling, and ultrasonic data.

\begin{figure}[H] \centering
\centering
\includegraphics[width=\figwidth]{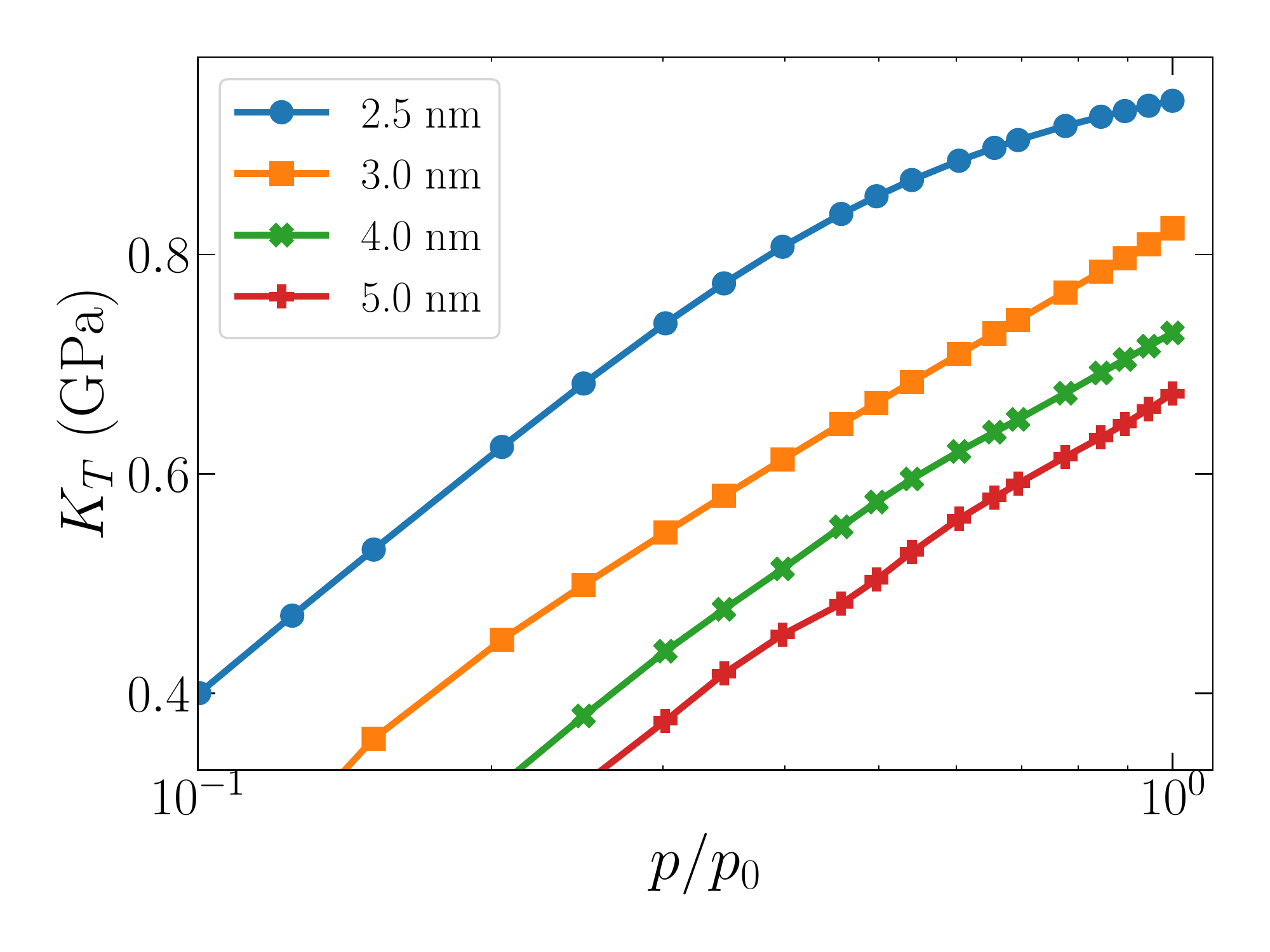} 
\caption{Isothermal bulk modulus $K_T$ of argon at 87.3 K confined in spherical pores of 2.5, 3, 4, and 5 nm in diameter as a function of relative pressure (calculated by GC-TMMC). Data from Ref.~\onlinecite{Gor2015compr}.}
\label{fig:Gor2015}
\end{figure}

\begin{figure}[H] \centering
\centering
\includegraphics[width=\figwidth]{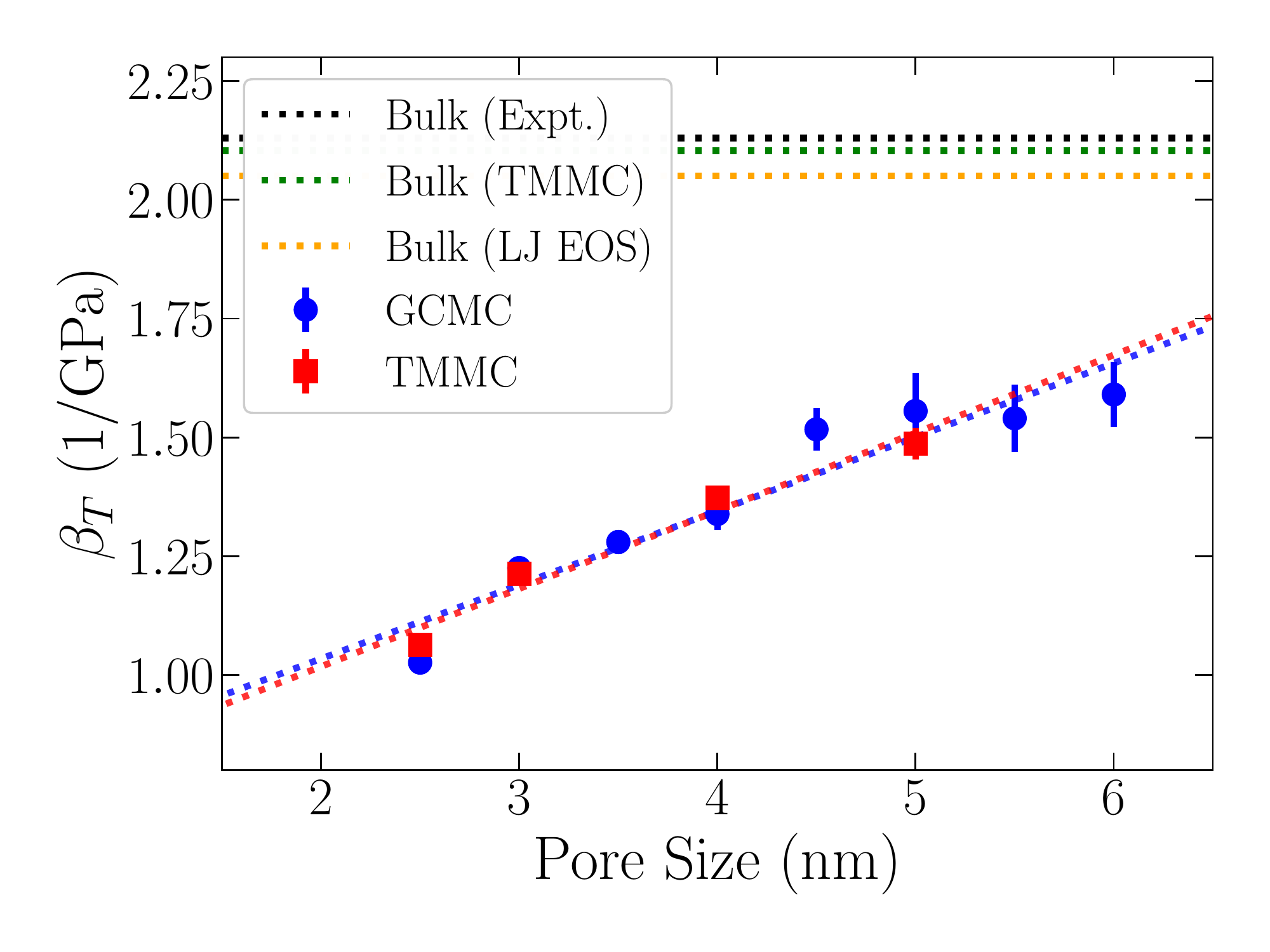}
\caption{Isothermal compressibility $\beta_T$ of liquid argon at 87.3 K at saturation ($p/p_0 = 1$) as a function of pore diameter. Horizontal dotted lines represent bulk argon based on experimental data, GC-TMMC simulations, and equation of state for LJ fluid~\cite{Johnson1993}. Compressibility is calculated using GCMC (circles) and GC-TMMC (squares). Diagonal dotted lines of corresponding marker color show linear fits for each method. Data from Ref.~\onlinecite{Gor2015compr}.}
\label{fig:Gor2015size}
\end{figure}

The next step was application of the grand canonical Monte Carlo simulations (including TMMC -- transition matrix Monte Carlo \cite{Siderius2013}) and the fluctuation formula Eq.~\ref{beta-fluct1} for the same system -- argon confined in silica pores~\cite{Gor2015compr}. Again, these simulations confirmed the experimentally-observed logarithmic dependence of the elastic modulus on the vapor pressure (i.e., compare Fig.~\ref{fig:Schappert2014} and Fig. \ref{fig:Gor2015}). Additionally, by varying the pore sizes from 2.5 to 6 nm in GCMC simulations, Gor et al.~\cite{Gor2015compr} found that the compressibility at $p = p_0$ is significantly lowered by confinement and is much lower for the smaller pore sizes and proposed a linear dependence of compressibility on the pore diameter (Fig. \ref{fig:Gor2015size}). However, additional calculations for larger pore sizes~\cite{Gor2017Biot, Dobrzanski2018} suggested a different dependence: a linear relation between the reciprocal values -- isothermal modulus ($K = \beta_T^{-1}$) versus reciprocal pore size $d^{-1}$. Finally, Gor and Gurevich~\cite{Gor2018Gassmann} re-analyzed the Schappert and Pelster experimental data~\cite{Schappert2014} using the Gassmann equation, and demonstrated quantitative agreement with the GCMC simulations data from Refs.~\onlinecite{Gor2015compr, Dobrzanski2018}. 

\begin{figure}[H] \centering
\centering
\includegraphics[width=\figwidth]{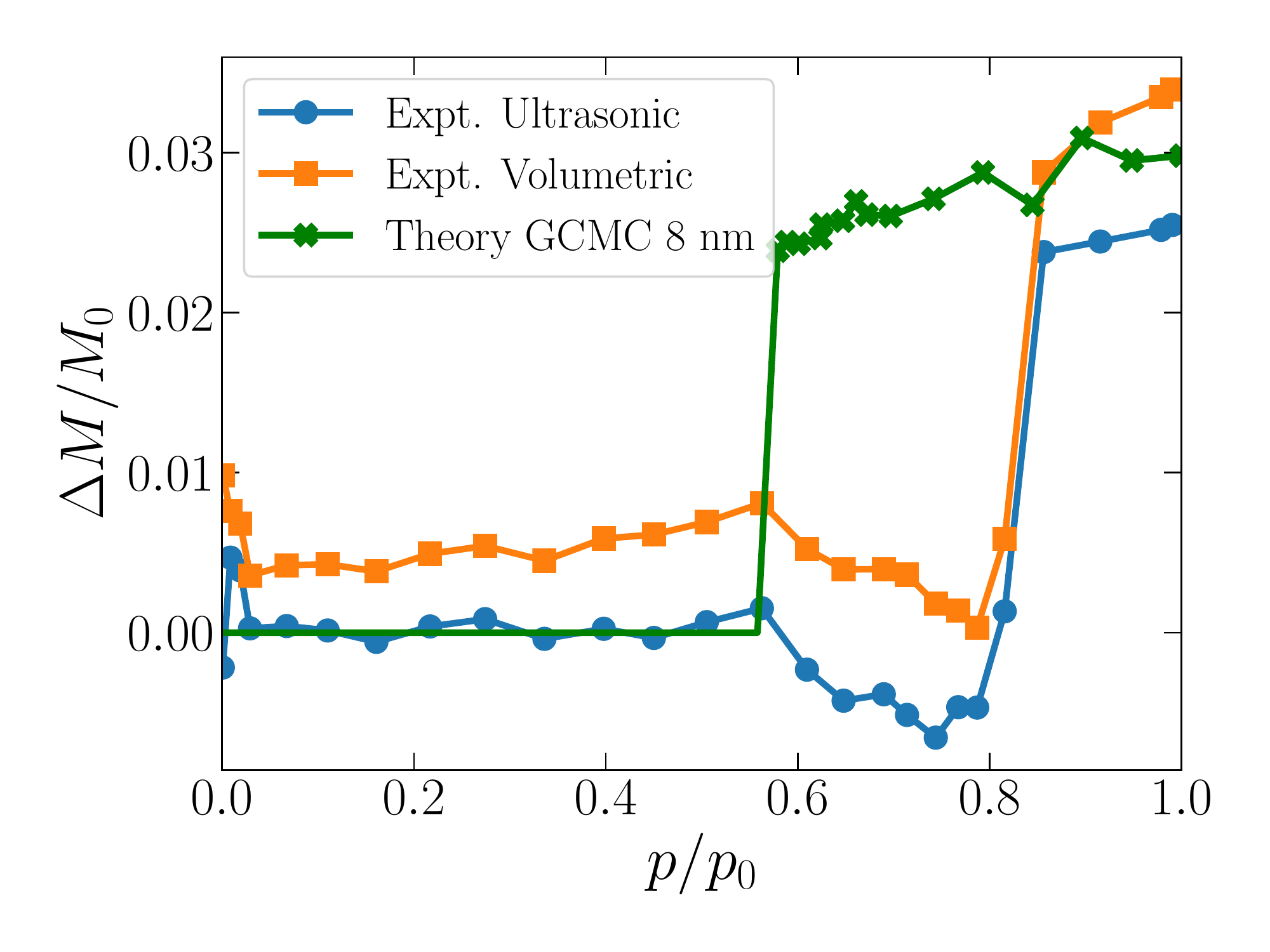}
\caption{Relative change of the longitudinal modulus of porous sample during nitrogen adsorption as a function of relative vapor pressure. The experimental ultrasonic curve are calculations based on speed of ultrasonic waves, the experimental volumetric curve is calculated from the combination of ultrasonic data for the longitudinal waves and volumetric data for the mass change, and the theoretical calculation is based on fluctuation of nitrogen molecules during GCMC simulations in a 8 nm spherical pore. Data from Refs.~\onlinecite{Maximov2018, Maximov2020}.}
\label{fig:Maximov2018}
\end{figure}

\begin{figure}[H] \centering
\centering
\includegraphics[width=\figwidth]{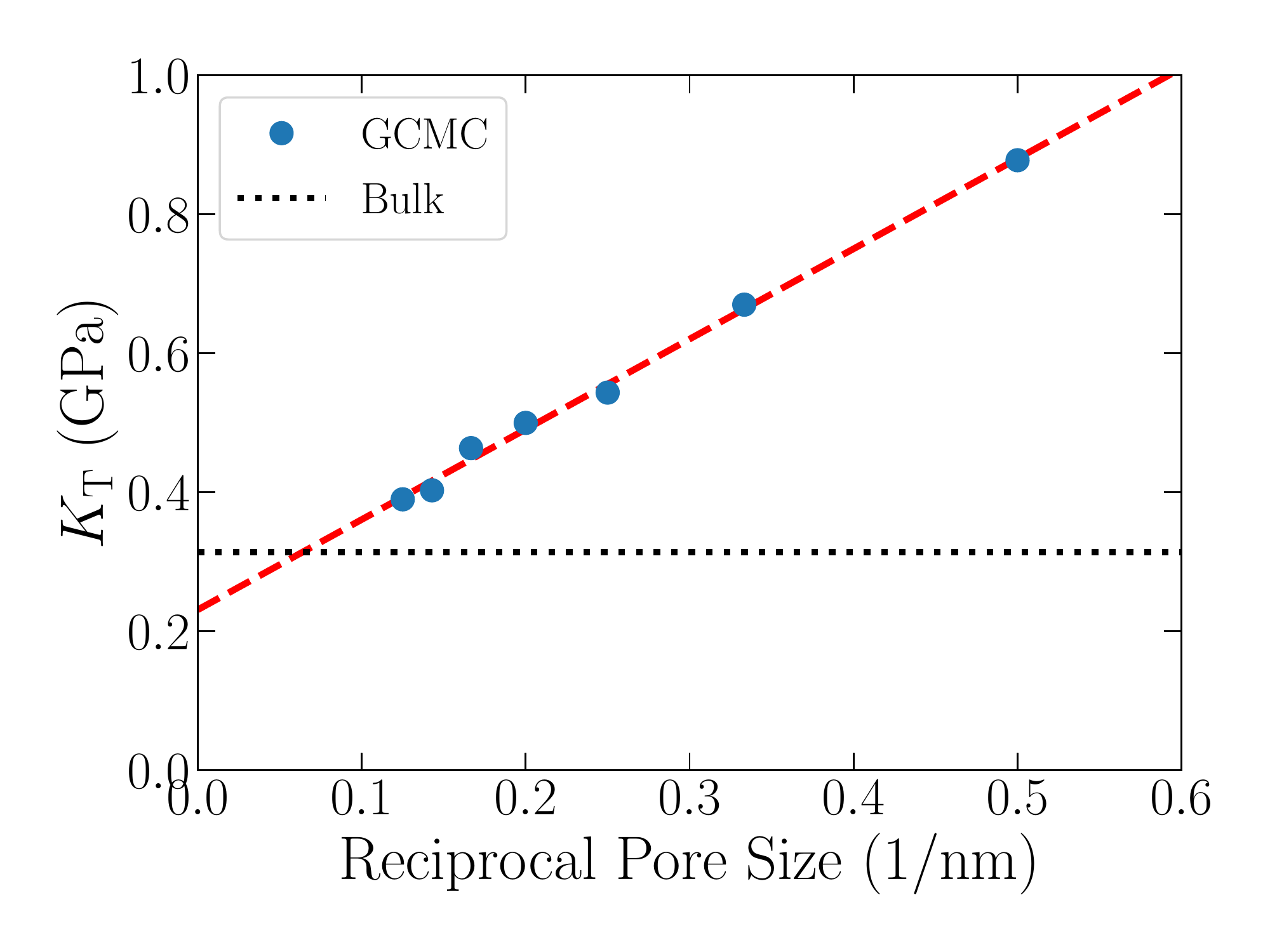}
\caption{Bulk elastic modulus at saturation ($p/p_0 = 1$) for nitrogen at 77 K in pores of various sizes, calculated from GCMC simulations along with a linear fit (red dashed line). The horizontal dotted line represents the elastic modulus of bulk liquid nitrogen at 77 K. Data from Refs.~\onlinecite{Maximov2018, Maximov2020}.}
\label{fig:Maximov2018size}
\end{figure}

Another work that directly compared GCMC data for the elastic modulus to ultrasonic experiments was done by Maximov and Gor for the system of nitrogen adsorption in nanopores~\cite{Maximov2018}. They calculated the isothermal elastic modulus of confined liquid nitrogen from molecular simulations, and also used the ultrasonic data from Warner and Beamish \cite{Warner1988} to calculate the longitudinal and shear moduli of the sample as a function of vapor pressure. They showed that the nitrogen modulus predicted from Monte Carlo simulation, when plugged into the Gassmann Eq.~\ref{Gassmann}, matches well with the modulus calculated from the experimental data of Warner and Beamish. Figure \ref{fig:Maximov2018} shows the experimental data for the modulus of the Vycor glass sample filled with liquid nitrogen, as a function of the relative pressure of nitrogen. The modulus is calculated in two different ways: 1 -- when the mass change is measured from the volumetric adsorption data, and 2 -- when the mass change is determined from the change of the shear modulus \cite{Warner1988}. Although the two methods are quite different, the results are comparable. The theoretical curve is calculated based on the molecular modeling combined with application of the Gassmann equation. The results of which ends up close to the experimental data sets.  Furthermore, Ref.~\onlinecite{Maximov2018} showed that the elastic modulus calculated from confined nitrogen in a range of pore sizes provides a linear trend as a function of the reciprocal pore size $d^{-1}$, see Figure \ref{fig:Maximov2018size} in contrast to Fig.~\ref{fig:Gor2015size}.

Wave propagation in fluid-saturated porous media has been studied within the theoretical framework of poromechanics, starting from the pioneering works by Biot \cite{Biot1956i, Biot1956ii}, and many contributions by Coussy \cite{Coussy1987, Coussy2004}. Later works by Coussy \cite{Vandamme2010, Coussy2011}, as well as by Ba\v{z}ant \cite{Nguyen2020} included extension of poroelasticity to nanoporous media, in particular taking into account the effects of adsorption. However, the change of compressibility of fluids as a result of confinement, and its effects on wave propagation have not been discussed in the poromechanics literature.

Ultrasonic experiments, discussed in Sec.~\ref{sec:Expt} can probe the \textit{average} elastic properties of the confined fluids, but not the \textit{local} properties discussed in Section~\ref{sec:Local}. The microscopic structure and local properties of confined fluids can be probed by experiments based on neutron or X-ray scattering~\cite{Melnichenko2015, Morineau2020chapter}. To our knowledge those have not been applied for probing the elastic properties, except for the work of Nyg{\aa}rd et al., who employed X-ray scattering for probing the local compressibility of confined \textit{colloidal} fluid \cite{Nygaard2016, Nygaard2016Opinion}. Their results confirmed the theoretical predictions on local compressibility changes at the solvophobic interfaces~\cite{Bratko2007, Sarupria2009, Evans2015}, thus justifying the theories based on local properties calculations.

\section{Summary and Outlook}
\label{sec:Summary}

When fluids are confined in nanopores, many of their properties change compared to the same fluid in the bulk including the density, freezing point, transport coefficients, thermal expansion coefficient, etc. The presented review shows that the elastic properties of the confined fluid also differ from the fluid in the bulk. We summarized the works showing experimental evidence of the effects of confinement on the elastic moduli. However, the number of experimental studies reporting the elastic properties of confined fluids is limited; there is a demand for more experimental measurements which could explore the broad spectrum of the potential porous solid-fluid systems. To our knowledge, the experiments that probe the compressibility of confined fluids have been performed almost exclusively on samples of Vycor glass \cite{Warner1988, Page1995, Schappert2014}. Future experiments should focus on other nanoporous solids that have different pore sizes, pore shapes, surface properties, etc. In particular, a series of measurements on similar samples with different pore sizes could help verifying the pore size dependence of the elastic modulus predicted by molecular simulations \cite{Gor2014, Dobrzanski2018, Corrente2020}. Furthermore, ultrasonic experiments with a broader family of liquids are desired in order to explore how molecular properties, such as polarity, molecule size, and shape, affect the compressibility of confined fluids. Specifically, experiments are needed for fluids which have practical importance for geophysics, i.e., water, hydrocarbons, and carbon dioxide. The two latter compounds are of special interest at supercritical conditions, at which, according to molecular modeling, the compressibility is more sensitive to effects of confinement \cite{Corrente2020}.

The main theoretical results are the following: 
\begin{enumerate}
\item The dependence of the elastic modulus of confined fluid on the solvation pressure in the pore through the Tait-Murnaghan equation.~\cite{Keshavarzi2016, Gor2016Tait}
\item The linear dependence of the elastic modulus on the reciprocal pore size $1/d$.~\cite{Gor2014, Dobrzanski2018}
\item The effect of strength of solid-fluid interactions on the departure of compressibility from the bulk value.~\cite{Evans2015, Evans2015PRL, Gor2016Tait, Gor2017Biot}  
\item The consistency between the local and average elastic properties.~\cite{Sun2019density}
\item The applicability of the Gassmann equation to nanoporous media.~\cite{Gor2018Gassmann, Maximov2018}
\item Showing that multiple differing methods of molecular modeling (i.e., MD, GCMC, and DFT) and use of various thermodynamic ensembles are able to predict the same values for the elastic properties of the confined fluid.~\cite{Gor2017Biot, Evans2017, Sun2019density, Corrente2020}
\end{enumerate}
Note that these theoretical results have practical implications, in particular they suggest the pore-size dependent correction for parameters for the Gassmann equation often used by practitioners. The dependence of fluid compressibility on the pore surface properties could be important for processes such as enhanced oil recovery, or carbon dioxide sequestration, which cause the surface modifications of the geological porous media \cite{Espinoza2010, deLara2015}. 

Although the amount of theoretical works on compressibility of confined fluids is richer than experimental, there are open questions. This is because most of the molecular modeling studies reporting the compressibility of confined fluid present qualitative discussion, without a direct comparison and verification from experiments. Particularly, most of the theoretical predictions for compressibility of confined fluids focus on structureless models for molecules, without electrostatic interactions, often represented by the simple Lennard-Jones potential. Such models are only adequate for simple fluids such as argon, nitrogen, methane, etc. At the same time, confined fluids of practical interest include hydrocarbons of different chain lengths as well as water and brine. Simulation for systems, such as confined water \cite{Sarupria2009, Strekalova2012} or long-chain hydrocarbons \cite{Martini2010}, have been performed, but unlike for argon or nitrogen, the direct comparison to experimental data has not been done. Thus, combined experimental-theoretical studies for non-simple liquids, liquid mixtures (e.g., brine) and confined solid phases remain an open area. Such studies can be based on Monte Carlo or molecular dynamics simulations, or utilize the recent progress in development of classical DFT for modeling more complex liquids, including water~\cite{Emborsky2011, Jeanmairet2013, Jeanmairet2016, Sauer2017}.

Furthermore, even for those simple fluids, some of the questions remain unresolved: the calculation of the compressibility in the limits of the smallest and largest pores. In particular, the calculation of argon compressibility in micropores using grand canonical Monte Carlo simulation was hindered by numerical artifacts. On the other hand, calculation of compressibility in large mesopores, above $\SI{10}{nm}$, require prohibitively long computational time \cite{Dobrzanski2018}. Both limitations demand alternative methods for calculating the compressibility. The elastic properties of bulk fluids are limited to bulk modulus or compressibility, because the shear modulus of a fluid is zero. While some experimental observations suggest that it is also the case for confined fluids, other works report non-zero shear moduli. To our knowledge modeling works addressing this question are non-existent, suggesting another open problem.

Although molecular simulation is a powerful theoretical tool for predicting thermodynamic properties of fluids, its computational cost limits its practical application. Even for bulk fluids, engineering applications demand the use of equations of state. In the last two decades, numerous works published in the literature presented several attempts to develop equations of state for confined fluids. While typically those were not employed to predict compressibility, a recent work has demonstrated that it is feasible and showed that one of these equations~\cite{Travalloni2010} qualitatively predicts compressibility of confined fluids \cite{Dobrzanski2020}. Therefore, another open challenge is to adapt an existing equation of state, or develop a new one, which can provide quantitative predictions for compressibility of confined fluids.

Finally, a question on the relation between the elastic properties of the confined fluid predicted from thermodynamic theories using properties probed in ultrasonic experiments remains open. While a theory typically focuses on the calculation of the properties of the fluid alone, the experiments probe the fluid-saturated porous medium, i.e., the solid-fluid composite. Therefore, in order to compare the two, one needs to know the properties of the solid constituent and use an effective medium approximation applicable to calculate the composite properties. A recent work used some of the limited experimental data available in the literature to compare to theoretical calculations for the confined fluids properties; the comparison suggested that the classical Gassmann equation can serve as an adequate effective medium approximation \cite{Gor2018Gassmann}. However, a rigorous approach towards verifying it would require a direct simulation of the composite system -- such modeling has not been done before. Additionally, more experimental data (in particular beyond the Vycor glass) would be helpful for verifying the theories.

\section*{Acknowledgments}

We would like to acknowledge fruitful discussions with many different people at different times which helped shaping this review, in particular Partick Huber, Maxim Lebedev, Peter Monson, Rolf Pelster, Peter Ravikovitch, Daniel Siderius, John Valenza, and Rodolfo Venegas. G. Y. G. acknowledges the support by the National Science Foundation under Grant No. CBET-1944495. B. G. thanks sponsors of Curtin Reservoir Geophysics Consortium for their financial support, and the support from the Australian Research Council through project DP190103260.

\section*{Data Availability Statement}

Data sharing is not applicable to this article as no new data were created or analyzed in this study.

\end{document}